\documentclass[twocolumn,aps,pre,showpacs]{revtex4}
\usepackage{epsfig}

\begin{document}
\title{Coarsening in potential and nonpotential models of oblique 
stripe patterns}

\author{J. R. Gomez-Solano}

\affiliation{Instituto de F\'\i sica, Universidad Nacional Aut\'onoma 
de M\'exico, Apartado Postal 20-364, 01000 M\'exico, D.F., M\'exico}

\author{D. Boyer}
\email{boyer@fisica.unam.mx}
\affiliation{Instituto de F\'\i sica, Universidad Nacional Aut\'onoma 
de M\'exico, Apartado Postal 20-364, 01000 M\'exico, D.F., M\'exico}

\date{\today}

\begin{abstract}
We study the coarsening of two-dimensional oblique stripe patterns by 
numerically solving potential and nonpotential anisotropic Swift-Hohenberg 
equations. Close to onset, all models exhibit 
isotropic coarsening with a single characteristic length scale growing 
in time as $t^{1/2}$. Further from onset, the characteristic lengths 
along the preferred directions $\hat{x}$ and $\hat{y}$ grow with 
different exponents, close to $1/3$ and $1/2$, 
respectively. In this regime, one-dimensional dynamical scaling relations hold.
We draw an analogy between this problem and Model A
in a stationary, modulated external field.
For deep quenches, nonpotential effects produce a complicated
dislocation dynamics that can lead to either arrested 
or faster-than-power-law growth, depending on the model considered.
In the arrested case, small isolated domains shrink down to a finite size 
and fail to disappear. A comparison with available experimental results of
electroconvection in nematics is presented.
\end{abstract}
\pacs{47.54.-r, 47.55.P-, 61.72.Cc}
\maketitle

\section{Introduction}

Domain coarsening of uniform phases in systems locally
in thermal equilibrium has received a lot of attention 
over the past decades \cite{gunton,bray}. After quenching a system below a 
transition point, ordered regions of the lower symmetry phase form and
their characteristic length $L$ grows slowly, usually as a power 
law at large times, $L\sim t^n$. Numbers of theoretical schemes based on 
the dynamical scaling hypothesis and on the study of the relaxation of 
topological defects have been proposed to infer
coarsening exponents and address their 
universality \cite{bray,halperin,brayrutenberg,bray1,lifshitz,allen}.

Less understood is the phase
ordering kinetics in systems driven out-of-equilibrium, in particular
those forming dissipative structures like regular patterns. 
Basic non-equilibrium structures are the modulated phases (stripes) of well
defined periodicity that appear above an instability threshold in thermal 
convection of fluids, in driven chemical reactors or in electroconvection 
of nematic liquid crystals 
\cite{crosshohenberg,bowman,gollub,rabinovich}. 

A body of numerical results
\cite{elder1,elder2,crossmeiron,hou,christensen,boyervinals1,
taneike,boyervinals2,qian1,qian2,paul,galla,xu,yokojima} suggests that 
the ordering kinetics of stripe patterns following a quench is not 
universal but model-dependent, and even, parameter-dependent.
Modulated phases in isotropic systems actually exhibit a rich variety 
of topological defects (dislocations, disclinations, grain boundaries) 
not observed in uniform phases. There might exist many mechanisms 
controlling the coarsening of stripe patterns: 
for instance, the annihilation of disclination quadrupoles, as observed 
experimentally \cite{harrison1,harrison2,ruiz}
in block-copolymer melts with cylindrical mesophases (leading to $n=1/4$), 
or grain boundary motion in the Swift-Hohenberg model close to onset ($n=1/3$)
\cite{boyervinals1,boyervinals2,qian1,xu}. 

For a given system, the growing length scale may even be not unique and the
dynamical scaling hypothesis not fulfilled.
Different defect-based definitions of the size $L$ of the 
ordered striped domains can lead to different coarsening exponents
\cite{crossmeiron,christensen,qian2,paul}. 
This effect seems to be more pronounced far from the onset of stripe formation.
In this regime, the growth of order often becomes very slow or arrested 
\cite{hou}, in particular due to pattern-induced pinning effects acting on 
defects \cite{boyervinals1,boyervinals2,xu}.

In addition, unlike equilibrium phases, the dynamics of nonequilibrium 
patterns is {\it a priori} not driven by the minimization of a free-energy
functional but better modelled by nonpotential (nonrelaxational) 
equations \cite{crosshohenberg,gollub,rabinovich,manneville}. 
Recent simulations of the 
Boussinesq equations for Rayleigh-B\'enard convection \cite{paul} suggest 
that multiple length scales must be defined to characterize the coarsening 
of nonpotential stripe patterns. Adding nonpotential terms to potential models 
\cite{crossmeiron,huangvinals} (or, similarly, taking into account 
hydrodynamics in the order parameter description of block-copolymer 
melts \cite{yokojima}) can noticeably increase some coarsening rates
and lead to qualitative changes in the structure of the
defected patterns. 

In this context, systems forming anisotropic patterns have recently attracted 
a particular attention \cite{purvis,kamaga,griffith,boyer,mazenko}. 
An example are the oblique phases, composed of
stripes that make a fixed angle, $\theta$ or $-\theta$, with respect 
to some $\hat{x}$-axis. Domain coarsening in these systems have been
studied in details experimentally in electroconvection of 
nematics \cite{purvis,kamaga,griffith}.
Anisotropic stripes may be regarded as topologically simpler 
than isotropic ones because of their finite number of 
orientations. However, some challenging difficulties mentioned 
above for isotropic systems still remain, such as the presence
of various sorts of defects and nonpotential effects.

The experimental studies of refs. \cite{purvis,kamaga} on oblique stripes
report that chevron grain boundaries and dislocations arrays 
(the domain interfaces along the $\hat{x}$ and $\hat{y}$ directions, 
respectively) have very different mobilities. After a quench, their 
respective densities have different power-law decays with time, 
both with unusually small exponents.
These results remain largely unexplained theoretically.
A first numerical study \cite{boyer} based on a potential anisotropic 
Swift-Hohenberg equation for oblique stripes \cite{peschkramer} 
reproduced the decay of the dislocation density but predicted a much
faster ordering in the $y$ direction than observed experimentally.

We revisit this problem by introducing in Section II
nonpotential versions (never considered before, to our knowledge) 
of the anisotropic model used in \cite{boyer}. 
The wavenumbers selected in these models are comparable to the 
experimental ones. In Section III, we define an orientational order 
parameter, its correlation lengths along the preferred 
$\hat{x}$ and $\hat{y}$ directions, and the defect densities (chevron 
and dislocations). In Section IV, we find two robust coarsening regimes
at moderate quench depths, independent of the potential or nonpotential 
nature of the models: an isotropic
regime very close to onset and an anisotropic one (in partial agreement
with experiments) at intermediate quenches. One-dimensional dynamical 
scaling relations hold in the latter case.  
In section V, we show that, up to moderate quenches, the problem 
can be approximately reduced to a Model A in a stationary, spatially 
modulated external field. 
Far from onset, nonpotential effects can lead to arrested or very fast 
domain growth, depending on the model. In the arrested case, 
small domains only shrink down to a finite size. 
Conclusions are presented in Section VI.

\section{Basic models and their first order amplitude equations}

An electroconvection set-up consists of a doped nematic liquid crystal 
confined between two plates with planar alignment. When an ac transverse 
electric field is applied, normal or oblique stripes (among other patterns) 
can form depending on the frequency \cite{kramer,buka,dennin,funfschilling}.
Some time ago, Pesch and Kramer (PK) introduced a phenomenological 
Swift-Hohenberg-like model that exhibits a transition from normal to 
oblique rolls \cite{peschkramer}:
\begin{equation}\label{anipotSH}
	\partial_t \psi = r \psi-\zeta^4(\nabla^2+k_0^2)^2 
	\psi-\frac{c}{k_0^4} \partial_y^4 \psi + 	
	\frac{2\eta}{k_0^4}\partial_x^2\partial_y^2\psi+NL[\psi],
\end{equation}
where $\psi(\vec{r},t)$ is the local order parameter, 
$NL[\psi]=-\psi^3$, $c$ and $\eta$ are dimensionless anisotropy 
parameters, $k_0$ a characteristic wavenumber, $\zeta$ (set to 
$1/k_0$ in the following) the coherence 
length and $r$ the main control parameter. In electroconvection of 
nematics with planar alignment, $x$ is the coordinate in the direction 
parallel to the undistorted director. 

Linear stability analysis of the uniform state $\psi(\vec{r},t)=0$ 
of Eq.(\ref{anipotSH}) against small periodic perturbations 
$\delta \psi(\vec{r},t) = \delta \psi_0 \exp(i\vec{k} \cdot \vec{r}+ 
\sigma t)$ gives the dispersion relation
\begin{equation}
\label{linearrate}	\sigma(p,q)=r-\frac{1}{k_0^4}
[\zeta^4 k_0^4(k_0^2-p^2-q^2)^2+cq^4-2\eta p^2 q^2],
\end{equation}
where $\vec{k}=p\hat{x}+q\hat{y}$.
Maximizing (\ref{linearrate}) with respect to $p$ and $q$ and looking 
for the values of $r$ for which $\sigma = 0$, one finds that some modes 
of finite wavenumber become marginally unstable when the control 
parameter $r$ exceeds the critical values $r_c^{(n)}$ or $r_c^{(o)}$:
\begin{eqnarray}   
\textrm{normal stripes:} & r_c^{(n)} = 0, \{ p_c^2 = k_0^2, q_c^2 = 0 \},
   \label{modosnorm}\\    
   \textrm{oblique stripes:} & r_c^{(o)} =  
   \frac{-\eta^2}{c+2 \eta - \eta^2} < 0,   
   \{ p_c^2 = \frac{c+\eta}{c+2\eta-\eta^2} k_0^2,\nonumber\\
           &q_c^2 = \frac{\eta}{c+2\eta-\eta^2} k_0^2 \}, \label{modosoblic}
\end{eqnarray}
where we have assumed that $c>0$. The marginally unstable wavevectors 
in the oblique case make an angle $\theta$ or $-\theta$, 
with respect to the $x$-axis, where $\theta = \arctan[\sqrt{\eta/(c+\eta)}]$.
These two-degenerate oblique modes, noted as $\vec{k}_c^{+}$ (zig) and 
 $\vec{k}_c^{-}$ (zag) in the 
following, are observed for $\eta > 0$ only. The model equation
(\ref{anipotSH}) actually reproduces the transition from normal to oblique 
stripes when $\eta$ is tuned from negative to positive values. 
 
In the following, we define the reduced control parameter as 
\begin{equation}\label{quench}	
\epsilon = r-r_c^{(o)}.
\end{equation}
Numerical solutions of Eq.(\ref{anipotSH}) with random  initial conditions 
for the field $\psi$ lead to the formation of oblique stripes if $c>0$ , 
$\eta>0$ and $r>0$ ({\it i.e.} $\epsilon>|r_c^{(0)}|$). Normal stripes 
were never observed close to onset for the parameter values used in this 
study. 
 
Domain coarsening in anisotropic oblique stripe patterns was recently 
studied with Eq. (\ref{anipotSH}) \cite{boyer}. This model has a 
potential structure, as it can be written as 
$\partial_t \psi = -\delta F/\delta \psi$, with $F$ a Lyapunov functional 
given by
\begin{eqnarray}\label{lyapanipotSH}	
F = \frac{1}{2k_0^4} \int &d\vec{r}&\left\{k_0^4(-r\psi^2+\psi^4/2)
+k_0^4 \zeta^4[(k_0^2+\nabla^2)\psi]^2\right. \nonumber\\	
&&\left.-2\eta (\partial_x \partial_y \psi)^2
+c(\partial_y^2 \psi)^2\right\}.
\end{eqnarray}
We propose here two nonrelaxational extensions of (\ref{anipotSH}) 
by replacing the $-\psi^3$ nonlinearity by terms that do not derive from
a functional. We consider the cases
\begin{eqnarray}	 
NL[\psi]&=&-\psi^3-\frac{c_1}{k_0^2}\psi (\nabla \psi)^2+
\frac{c_2}{k_0^2}\psi^2 \nabla^2 \psi,\	\textrm{Model I} \label{modelI}\\
NL[\psi]&=&	\frac{c_3}{k_0^4}\nabla^2 \psi (\nabla \psi)^2\nonumber\\
	&+&\frac{3-c_3}{k_0^2}(\partial_i \psi)(\partial_j \psi)	
	(\partial_i \partial_j \psi),\ \textrm{Model II} \label{modelII},
\end{eqnarray}
where $c_1,c_2,c_3$ are constants and $(i,j)=(x,y)$. 
These terms were previously considered in the isotropic 
Swift-Hohenberg model of Rayleigh-B\'enard convection \cite{greenside}
(see also \cite{crossmeiron}).
Expressions (\ref{modelI}) and (\ref{modelII}) are nonpotential when 
$c_1 \neq -c_2$ and $c_3 \neq 2$. We set $c_1 = c_2 = 1$ and 
$c_3 = 3$ in the following. 

As shown in the next Section, all the above models with random initial 
conditions lead to the formation of polycrystalline configurations 
as the one shown in Fig. \ref{fig:pattern}. Their large time evolution 
is controlled by the motion of grain boundaries separating domains of 
perfectly oriented zig and zag stripes. To describe patterns containing 
grain boundaries, we look for general solutions of the model equations 
of the form \cite{manneville,malomed}
\begin{equation}\label{ansatz}	
\psi(\vec{r},t) = A^+(\vec{r},t) e^{i \vec{k}_c^+ \cdot \vec{r}}
+A^-(\vec{r},t)e^{i \vec{k}_c^- \cdot \vec{r}}+c.c.,
\end{equation}
where $\vec{k}_c^{\pm} = p_c \hat{x} \pm q_c \hat{y}$, with $p_c$ and $q_c$ 
positive and given by Eq.(\ref{modosoblic}); $c.c.$ means the complex 
conjugate. Solutions of the form (\ref{ansatz}) describe polycrystalline
states of zig and zag domains with wavevectors close to the marginal ones. 
$A^+ \neq 0$ 
and $A^-\simeq 0$ within a zig domain, and vice-versa. The amplitudes 
$A^+$ and $A^-$ are both non-vanishing in the vicinity of 
a grain boundary.

In the limit $\epsilon\ll1$, 
the multiple scale formalism \cite{manneville,peschkramer}
can be applied to the three models (\ref{anipotSH}), (\ref{modelI}) and 
(\ref{modelII}). We find that, at first order in $\epsilon^{1/2}$, 
the dynamics of $A^+$ and $A^-$ is described by two coupled 
Ginzburg-Landau equations
\begin{eqnarray}\label{aniginzburg1}	
\partial_t A^+ &=&\epsilon A^+ 	+ \frac{4}{k_0^4}[p_c^2 \partial_x^2	
+2(1-\eta)p_c q_c \partial_x \partial_y\\	&+&(1+c)q_c^2\partial_y^2]A^+
-3\gamma|A^+|^2A^+-6\gamma|A^-|^2A^+\nonumber
\end{eqnarray}
\begin{eqnarray}\label{aniginzburg2}	
\partial_t A^- &=& \epsilon A^- + \frac{4}{k_0^4}[p_c^2 \partial_x^2	
-2(1-\eta)p_c q_c \partial_x \partial_y\\	
&+&(1+c)q_c^2\partial_y^2]A^-\nonumber	
-3\gamma|A^-|^2A^--6\gamma|A^+|^2A^-,
\end{eqnarray}
where $\gamma$ is a constant parameter whose expression depends on the 
model considered,
\begin{displaymath}    
\gamma = \left\{
\begin{array}{ll}        
 1, & {\rm potential\ PK\ model},\\	
 1+\frac{c_1+3c_2}{3}\left(\frac{c+2\eta}{c+2\eta-\eta^2}\right), 
 & {\rm Model\ I},\\	
 \left(\frac{c+2\eta}{c+2\eta-\eta^2}\right)^2, & {\rm Model\ II}.
\end{array} 
\right.
\end{displaymath}
Equations similar to (\ref{aniginzburg1})-(\ref{aniginzburg2}) 
were proposed as reduced models of oblique patterns 
in electroconvection \cite{bodenschatz,treiber} and thermoconvection 
\cite{plaut} of nematics with planar alignment close to onset. 
They can be recast as 
$\partial_t A^+ = -\delta F_{GL}/\delta \bar{A}^+$ and 
$\partial_t A^- = -\delta F_{GL}/\delta \bar{A}^-$, where $F_{GL}$ is a 
Lyapunov functional given by 
\begin{eqnarray}  
 F_{GL}&=&\int_{\ } d\vec{r}\left\{ -\epsilon 
 \left( |A^+|^2+|A^-|^2 \right)\right. \nonumber\\    
 &+&\frac{4}{k_0^4} \left| [p_c \partial_x+ (1-\eta)q_c \partial_y] A^
 +\right|^2   \nonumber\\       
 &+&\frac{4}{k_0^4}\left|[p_c \partial_x - (1-\eta)q_c]A^-\right|^2 \nonumber\\
     &+&\frac{4 \eta}{k_0^2}\left(|\partial_y A^+|^2+|\partial_y A^-|^2\right)
     \nonumber \\        
     &+& \left.\frac{3}{2}\gamma(|A^+|^4+|A^-|^4)+6\gamma |A^+|^2 |A^-|^2 
     \right\}, \label{fgl}
\end{eqnarray}
where $\bar{A}$ is the complex conjugate of $A$. Close to onset 
($\epsilon \ll 1$) the amplitude equations of anisotropic oblique striped 
domains are identical for the three models (\ref{anipotSH}), (\ref{modelI}), 
(\ref{modelII}) up to the multiplicative constant $\gamma$ 
(that can be absorbed in amplitude, time and space scales). The 
amplitude equations have 
a potential structure even when deriving from a nonpotential model. We 
therefore expect that very close to onset oblique stripe patterns exhibit a 
generic ordering dynamics, independently of the potential or nonpotential 
nature of the system. Amplitude equations (\ref{aniginzburg1}) and 
(\ref{aniginzburg2}) will be used to discuss some aspects of the 
coarsening process close to onset in Section \ref{sec:modelA}. For a more 
detailed description of defect dynamics far from onset, higher order 
contributions ($O(\epsilon^2)$) would be needed in the derivation of the 
amplitude and phase equations \cite{crosshohenberg,manneville}.

\section{Numerical method}

Equation (\ref{anipotSH}) with the nonlinear terms (\ref{modelI}) or 
(\ref{modelII}) is numerically solved 
by using a semi-implicit pseudospectral method and a time integration 
procedure described in \cite{chaos}. The two-dimensional space is 
discretized on a square grid of mesh size $\Delta x = 1$ with 
1024$\times$1024 nodes. The characteristic wavelength is set to 
$\lambda_0 = 2\pi/k_0 = 8\Delta x$. The time 
integration scheme is stable for large values of the time step 
$\Delta t$. We set $\Delta t = 0.5$ for all the values of the 
quench depth studied. The initial condition for $\psi$ is a random 
field with Gaussian distribution, of mean $\langle \psi \rangle = 0$ 
and variance $\langle \psi^2 \rangle = \epsilon$. Numerical results 
are averaged over 10 different initial conditions for each model and 
value of $\epsilon$.

\begin{figure*}[htp]          
\includegraphics[width=.4\textwidth]{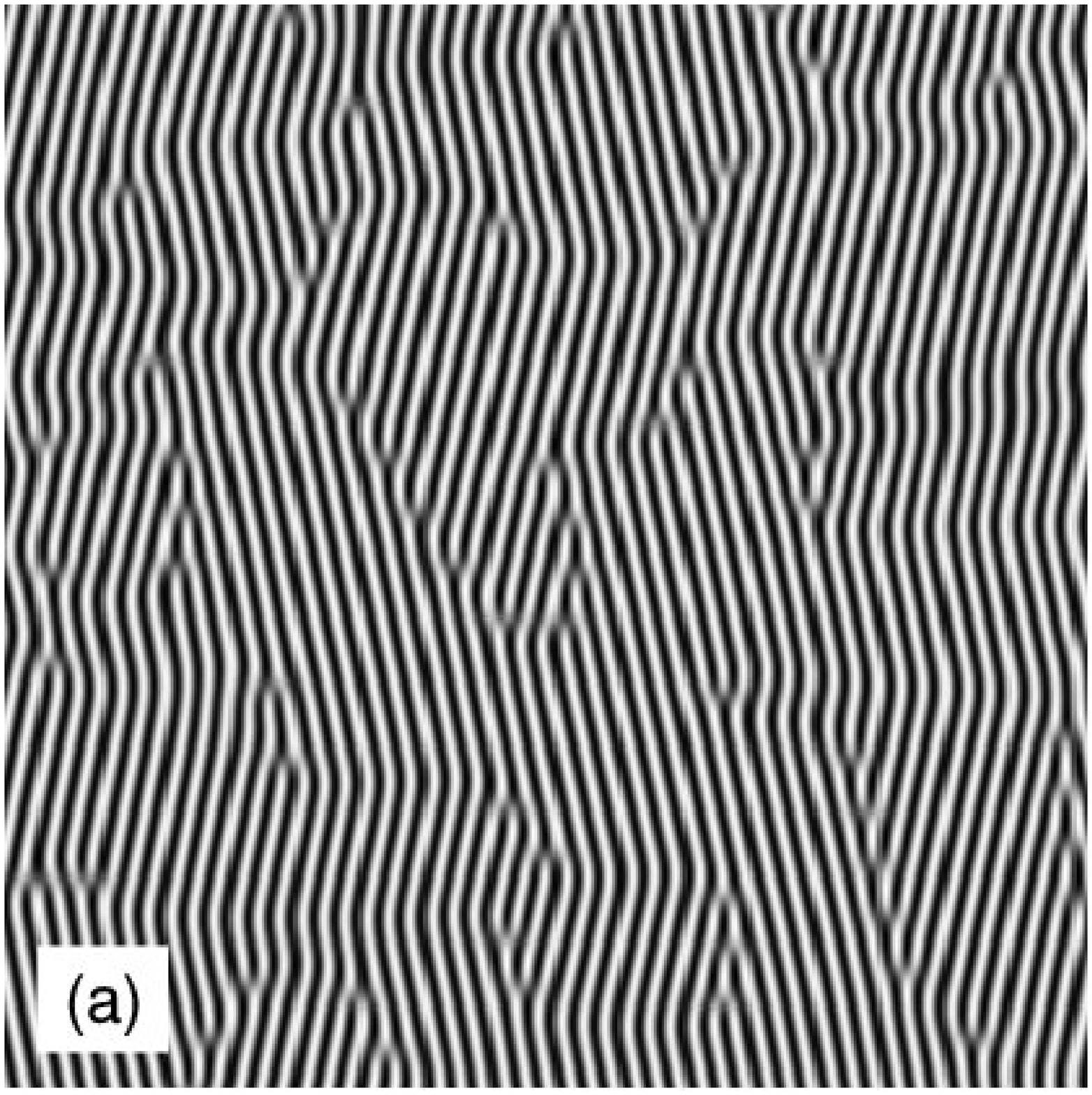}
 \hspace{.0in}\includegraphics[width=.4\textwidth]{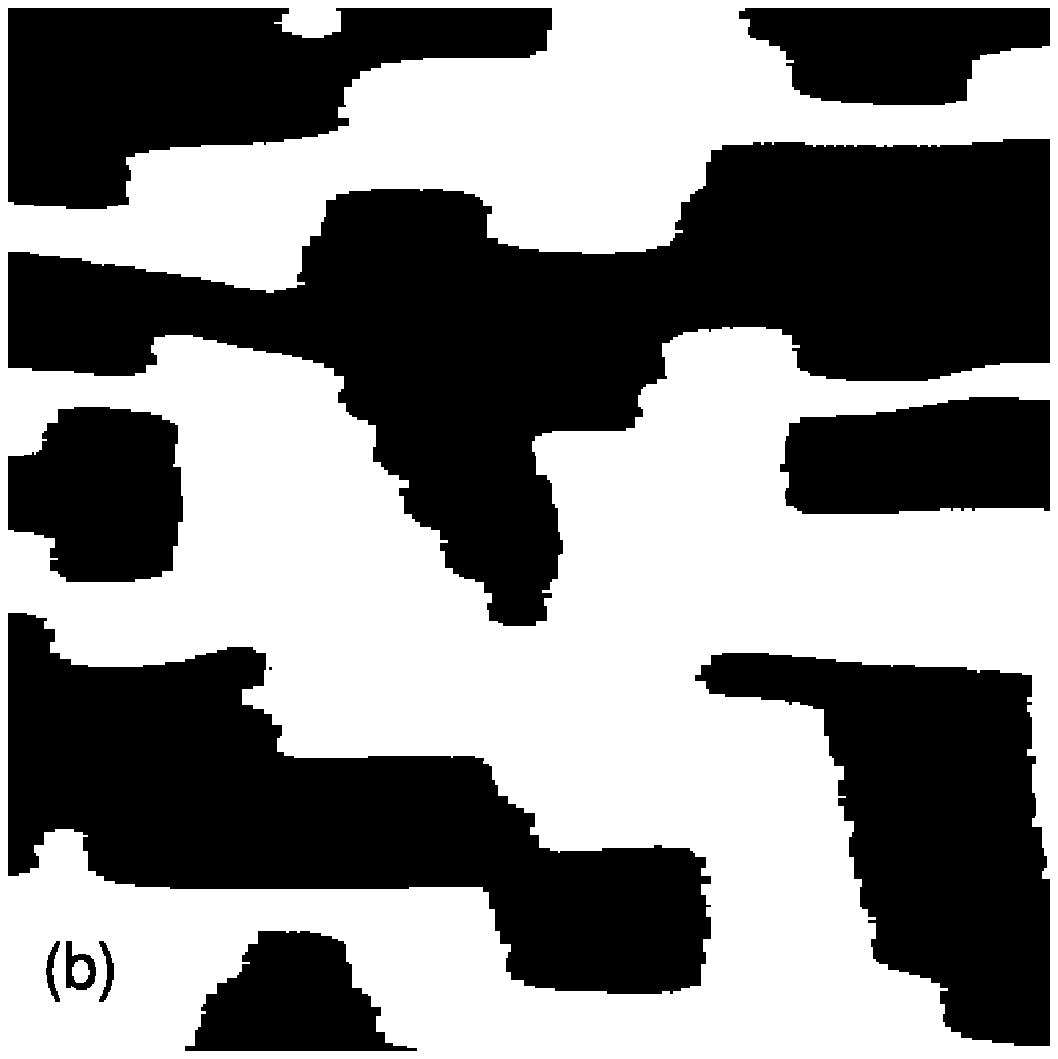}\\
 \vspace{.0in}         
 \includegraphics[width=.4\textwidth]{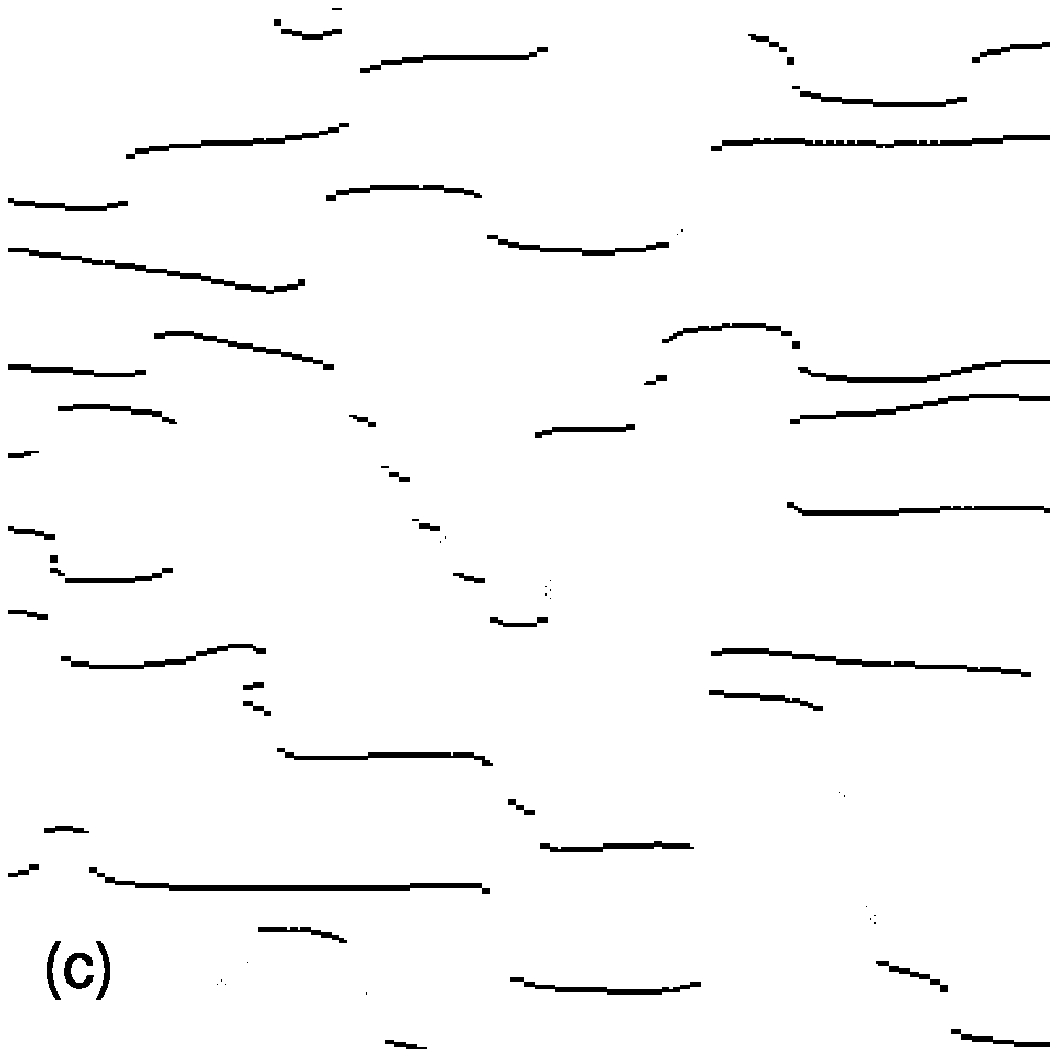}     \hspace{.0in}         
 \includegraphics[width=.4\textwidth]{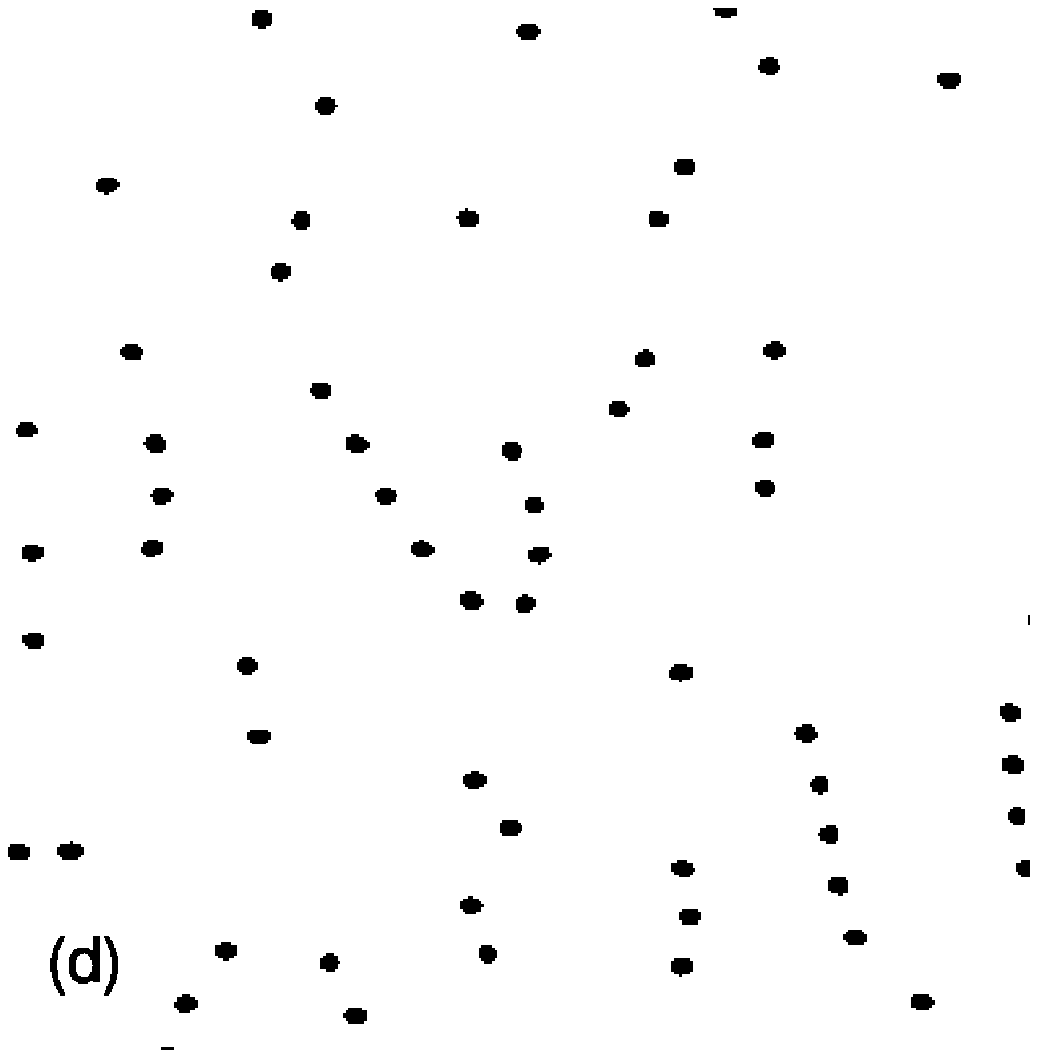}     
 \caption{\label{fig:pattern} a) Oblique stripe pattern at $t=500$ obtained 
 by numerically solving Eq. (\ref{modelI}) (Model I) with $\epsilon$ = 0.4118, 
 $c=12$, $\eta = \textrm{0.5}$ and    $c_1 = c_2 = 1$; 
 b) domain configuration of the stripe pattern in (a); c) chevron boundaries 
 of (a); d) dislocations of (a). We show 
	a fraction of $400 \times 400$ 
 nodes from a total discretized system of size $1024 \times 1024$.}
 \end{figure*}
Given relation (\ref{quench}) and the fact that a random initial Gaussian 
 field evolve towards oblique patterns only if $r>0$, we choose the parameters 
 in Eq.(\ref{anipotSH}) such that the smallest possible quench 
 ($\epsilon_{min}=-r_c^{(o)}$) is small compared with unity. We set $c=12$ 
 and $\eta =$ 0.5 in all the calculations, corresponding to 
 $-r_c^{(o)} = 0.0196$ and $\theta = \textrm{11.31}^{\circ}$ 
 ($\vec{k}_c^{\pm}/k_0=\textrm{0.9901} \hat{x} \pm \textrm{0.1980} 
 \hat{y}$).

 A typical configuration of oblique stripe domains is displayed in 
 Figure \ref{fig:pattern}(a) at $t = 500$ in the case of Model I for a 
 moderate quench. As observed in electroconvection experiments 
 \cite{purvis,kamaga} and previous numerical simulations of 
 (\ref{anipotSH}) \cite{boyer}, interfaces separating zig and zag 
 stripes are mainly of two kinds: horizontal chevron boundaries 
 where stripes change their orientation from $\theta$ to $-\theta$ 
 without any phase singularity, and slightly curved boundaries 
 composed of arrays of dislocations with an orientation close to the vertical
 direction. Close to onset, isolated 
 dislocations are scarcely observed within domains and do not seem to 
 control the ordering dynamics. Note that isolated dislocations 
 in anisotropic patterns are analogous to superfluid vortices, that 
 are known to have a diverging energy \cite{nelson}. This is due to 
 the fact that, unlike in isotropic stripes, spatial derivatives in 
 the first order amplitude equations 
 (\ref{aniginzburg1}-\ref{aniginzburg2}) can be recast into 
 a Laplacian form after an axis rotation.

Zig and zag domains can be identified by means of a (scalar) 
orientational local order parameter, $\psi_d$, defined as 
\[
    \psi_d(\vec{r},t)=\left\{    
    \begin{array}{ll} +1, 
    & \mbox{if $(\partial_x \psi)(\partial_y \psi) \ge 0$ (zig)},\\
      -1, & \mbox{if $(\partial_x \psi)(\partial_y \psi) < 0$ (zag)}.        
      \end{array}        
      \right.
      \]
The orientation field $\psi_d$ of the pattern of Fig. \ref{fig:pattern}(a) 
is shown in Fig. \ref{fig:pattern}(b). 

Due to the anisotropy of the problem, we measure the growth of order along 
the $\hat{x}$ and $\hat{y}$ directions separately. For this purpose we 
introduce an orientational structure factor, defined as
\begin{equation}\label{sfpsid}	
S(\vec{q},t) \equiv \langle \tilde{\psi}_d(\vec{q},t) 
\tilde{\psi}_d(-\vec{q},t) \rangle,
\end{equation}
where $\tilde{\psi}_d$ is the 2D Fourier transform of $\psi_d$ and
the brackets indicate averages over different initial conditions.
We define the one-dimensional structure factors along $\hat{x}$ and $\hat{y}$ 
as 
\begin{equation}\label{sx}   
S_x(q_x,t) = \int_{-\infty}^{+\infty} S(\vec{q},t) dq_y,
\end{equation}
and
\begin{equation}\label{sy}
    S_y(q_y,t) = \int_{-\infty}^{+\infty} S(\vec{q},t) dq_x,
\end{equation} 
respectively. It can easily be shown that (\ref{sx}) 
and (\ref{sy}) are the 1D Fourier transforms of the equal time correlation 
functions along the directions $\hat{x}$ and $\hat{y}$, respectively,
\begin{equation}\label{cx}    
C_x(x,t)= \int \langle 
\psi_d(\vec{r}',t) \psi_d(\vec{r}'    +x\hat{x},t) \rangle d\vec{r}',
\end{equation}
\begin{equation}\label{cy}
    C_y(y,t)= \int \langle \psi_d(\vec{r}',t) \psi_d(\vec{r}'   
    +y\hat{y},t) \rangle d\vec{r}'.\end{equation}
    We define the correlation lengths in the $\hat{x}$ and $\hat{y}$
     directions from the inverse width of the curves $S_x(q_x,t)$ 
     and $S_y(q_y,t)$ in Fourier space:
     \begin{equation}\label{lx}    L_x(t)
     =\left( \frac{\int_{-\infty}^{\infty} |q_x| S_x(q_x,t) dq_x}    
     {\int_{-\infty}^{\infty} S_x(q_x,t) dq_x} \right)^{-1},
     \end{equation}
     \begin{equation}\label{ly}    
     L_y(t)=\left( \frac{\int_{-\infty}^{\infty} |q_y| S_y(q_y,t) dq_y}    
     {\int_{-\infty}^{\infty} S_y(q_y,t) dq_y} \right)^{-1}. 
     \end{equation}

Like in isotropic systems, the defect densities should 
be related the correlation lengths defined above if dynamical 
scaling holds. In the case of 
oblique stripe domains, we investigate separately the evolution of 
of chevron ($\rho_{ch}$) and dislocation ($\rho_{dis}$) densities. 
Chevron boundaries are identified as the lattice nodes 
$(x',y')$ where
\begin{equation}\label{cheviden}    
|\psi_d(x',y'+1,t)- \psi_d(x',y',t)|=2,
\end{equation}
The chevron density is defined as the fraction of such nodes, marked in black 
in Fig. \ref{fig:pattern}(c). We do not consider as belonging to 
a chevron boundary isolated horizontal clusters of nodes 
satisfying (\ref{cheviden}) shorter than half the stripe period.
Dislocations are identified by using the 
filtering procedure described in \cite{hou}. A typical configuration 
of dislocations is shown in Fig. \ref{fig:pattern}(d), corresponding 
to the pattern of Fig. \ref{fig:pattern}(a). We define 
numerically $\rho_{dis}$ as the fraction of area occupied by the black 
regions in Fig. \ref{fig:pattern}(d).

\section{Results}\label{sec:results}

\subsection{Wavenumber selection}

\begin{figure}[htp]          
\includegraphics[width=.45\textwidth]{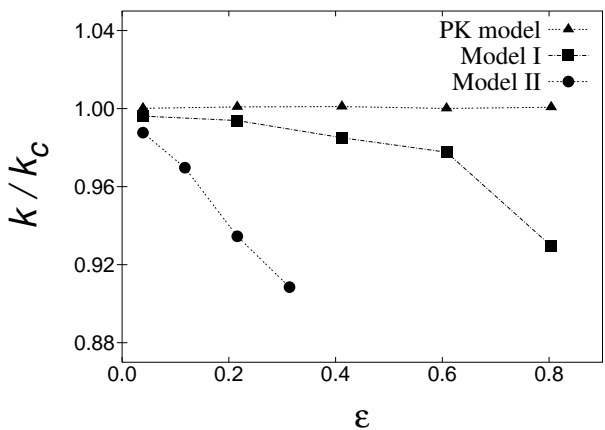}
 \caption{\label{fig:qmax} Selected wavenumber as a function of
 $\epsilon$ for the three models studied.}
 \end{figure}

Before investigating domain growth we address the issue of wavenumber 
selection at short times, as it can play an important role in the 
subsequent defect dynamics \cite{crossmeiron,funfschilling,paul,
huangvinals}. The wavenumber $k$ of the background stripes at 
time $t$ is numerically obtained by averaging the modulus of the four 
wavevectors where the order parameter structure factor 
$\langle\tilde{\psi}(\vec{q},t)\tilde{\psi}(-\vec{q},t)\rangle$ is maximum
($\tilde{\psi}$ being the Fourier transform of $\psi$). 

Like for isotropic potential equations \cite{crosshohenberg,bowman},
the selected wavenumber $k$ in the PK model (\ref{anipotSH}) is the one that 
minimizes the Lyapunov functional (\ref{lyapanipotSH}). 
After a short transient $k$ reaches an asymptotic value 
that is independent of $\epsilon$, as shown in Fig. \ref{fig:qmax},
and given by the marginal wavenumber maximizing the growth
rate (\ref{linearrate})
\begin{equation}\label{marginalwn}	
k_c = \sqrt{p_c^2+q_c^2}= \sqrt{\frac{c+2\eta}{c+2\eta-\eta^2}}k_0.
\end{equation}
Nonpotential models of isotropic patterns generally exhibit 
nontrivial wavenumber selection far from onset, with $k < k_c$.
The value of $k$ depends on $\epsilon$ and is not determined by any 
minimization principle \cite{crossmeiron,paul,hari,huangvinals,tesauro}. 
We observe similar behavior for the nonpotential anisotropic models. 
The selected wavenumber in the nonpotential models I and II is very close 
to $k_c$ sufficiently close to onset.
As $\epsilon$ is increased, the differences between the three models 
become more pronounced. For the nonpotential models I and II, $k$ 
decreases with increasing $\epsilon$ (see Fig. \ref{fig:qmax}), 
in qualitative agreement with electroconvection 
experiments far from onset and below the Lifshitz point 
\cite{funfschilling}. This aspect further justifies the use 
of nonpotential equations for modeling electroconvection.
The dependence of $k$ on $\epsilon$ is weaker for Model I than for 
Model II, though. In Model I, $k$ remains close to the marginal value 
($|k-k_c|/k_c < 0.03$) even for quench depths as large as 
$\epsilon \approx \textrm{0.6}$.

\subsection{Correlation lengths}\label{sec:correl}

\begin{figure}[htp]        
\includegraphics[width=.45\textwidth]{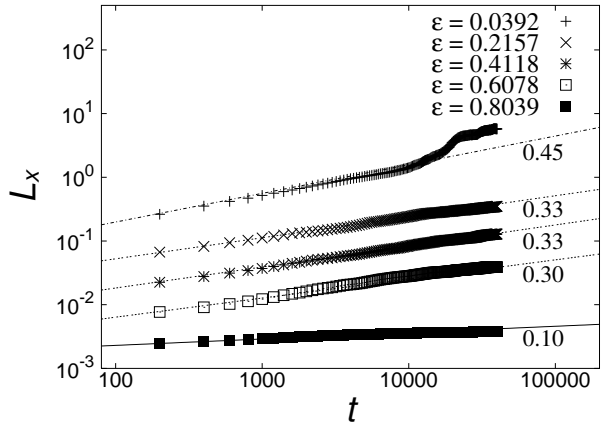}\\     
\vspace{.2in}
\includegraphics[width=.45\textwidth]{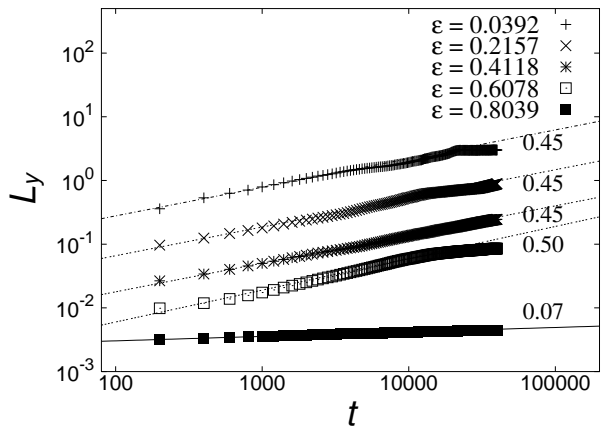}    
\caption{\label{fig:m1correl}Correlation lengths along $\hat{x}$ and $\hat{y}$ 
vs. time for Model I at different values of $\epsilon$. 
The curves have been displaced from their original position for
clarity.}
\end{figure}

\begin{figure}[htp]     
\includegraphics[width=.45\textwidth]{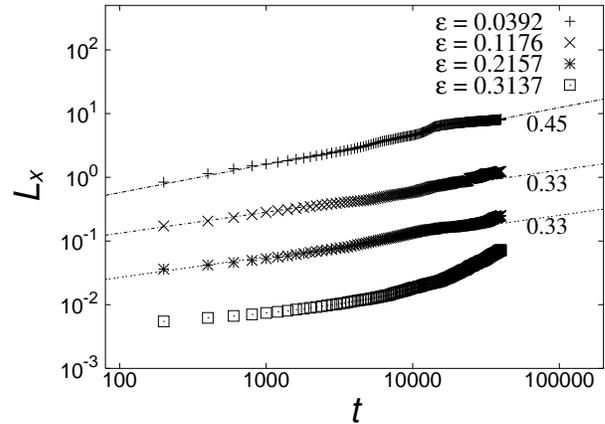}\\     
\vspace{.2in} 
\includegraphics[width=.45\textwidth]{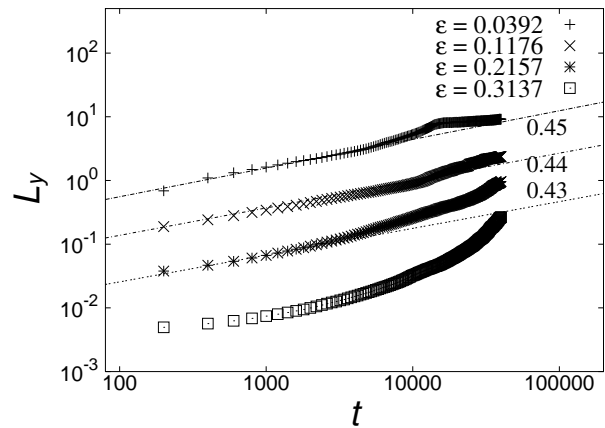} 
\caption{\label{fig:m2correl}Correlation lengths along $\hat{x}$ and 
$\hat{y}$ vs. time for Model II at different values of 
$\epsilon$.}
\end{figure} 

The time evolutions of the correlation lengths $L_x(t)$ and $L_y(t)$ are 
shown in Figs. \ref{fig:m1correl} and \ref{fig:m2correl} for 
different values of $\epsilon$ and for each nonpotential model. 

Close to onset ($\epsilon = \textrm{0.0392}$), the coarsening rates are 
well 
fitted by the power laws
\begin{equation}\label{eq:correllaw}
	L_x(t) \sim t^{1/z_x}, \qquad L_y(t) \sim t^{1/z_y}.
\end{equation}
	For both Models I and II, we find 
	$1/z_x = 1/z_y \approx \textrm{0.45}$  in the intermediate time 
	regime. Finite size effects become important at large times. 
	These results are similar to those obtained for the PK 
	potential model \cite{boyer} where exponents close to $1/2$, 
	the value typical of curvature driven growth, were reported 
	near onset. In Section \ref{sec:modelA} below, we present 
	analytical arguments showing that the three models 
	proposed can be approximatively reduced, close to onset 
	only, to Model A for a non-conserved order parameter \cite{halperin}.

At intermediate values of $\epsilon$, comprised at least in the 
interval $[0.2,0.4]$ in the case of Model I, the growth of both 
correlation lengths is well fitted by power laws with two distinct 
exponents: $1/z_x \approx 0.33$ and $1/z_y \approx \textrm{0.45}$, 
as shown in Fig. \ref{fig:m1correl}. 
Similar results were obtained for the potential PK model 
\cite{boyer}. These features are also observed in Model II, but in a 
smaller interval of values of $\epsilon$. For  $\epsilon=0.1176$, the 
scaling laws $L_x\sim t^{0.33}$ and $L_y\sim t^{0.45}$ hold until 
$t\simeq 10^4$, see Fig. \ref{fig:m2correl}. 
\begin{figure}[htp]      
\includegraphics[width=.45\textwidth]{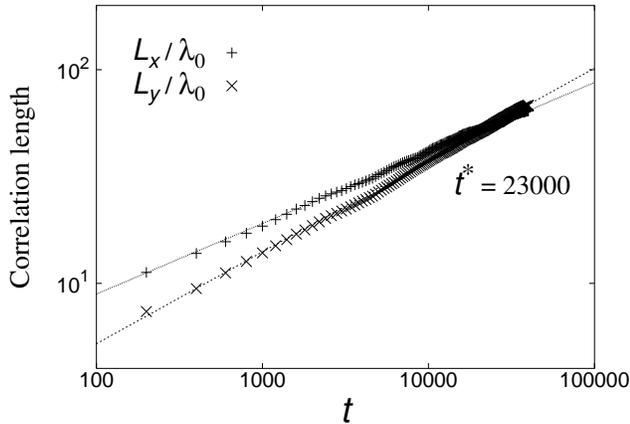}    
\caption{\label{fig:lxly}Time evolution of the correlation lengths along 
$\hat{x}$ and $\hat{y}$ for Model I with $\epsilon = \textrm{0.4118}$. 
The characteristic time such that $L_x = L_y$ is $t^* \approx 23000$.}
\end{figure}
This two-exponents regime observed at intermediate quenches in all
models indicates that $L_y$ grows faster than $L_x$. However, as shown 
in Fig. \ref{fig:lxly} where both lengths are plotted together for 
Model I, domains are in average longer along $\hat{x}$ than along 
$\hat{y}$ at short times. Therefore, there is a characteristic time 
$t^*$ such that $L_x(t^*) = L_y(t^*)$. When $t>t^*$, domains become 
longer along $\hat{y}$ than $\hat{x}$. A qualitatively similar behavior  
is reported in electroconvection experiments performed in \cite{purvis}, 
where domains are more elongated along $\hat{x}$ than along $\hat{y}$ 
at short times, although the growth along $\hat{y}$ is faster.

For deep quenches, the growth rates of $L_x(t)$ and $L_y(t)$ 
become extremely slow in Model I, as shown in Fig. \ref{fig:m1correl} for 
$\epsilon = \textrm{0.8039}$. 
The transition to such arrested dynamics is relatively abrupt, since 
for $\epsilon = \textrm{0.6078}$, $L_y$ and $L_y$ are still well 
described by the laws $t^{1/2}$ and $t^{1/3}$, respectively.  
Similar frozen states observed in the potential PK model
were explained by the presence pinning effects acting on 
defects at large quenches \cite{boyer}.
However, as discussed in Section \ref{sec:nonrel}, freezing probably has
a different origin in the nonpotential case.

Whereas the ordering dynamics in Model I is similar to that of the 
potential case for a wide range of values of $\epsilon$, the 
nonpotential terms of Model II have dramatic effects at large $\epsilon$. 
For $\epsilon = \textrm{0.2157}$ and $\epsilon = \textrm{0.3137}$, 
the correlation lengths grow {\em faster} than a power law at intermediate 
and large times, see Fig. \ref{fig:m2correl}. Only at short times 
($t<10^3)$, an apparent scaling regime with $1/z_x \le 1/3$ and 
$1/z_y \le \textrm{0.45}$ can be observed. A secondary bifurcation 
from oblique to normal stripes actually occurs in Model II close to 
$\epsilon \approx \textrm{0.4}$, preventing the study of oblique domain 
coarsening for larger quenches.

\begin{figure}[htp] 
\includegraphics[width=.45\textwidth]{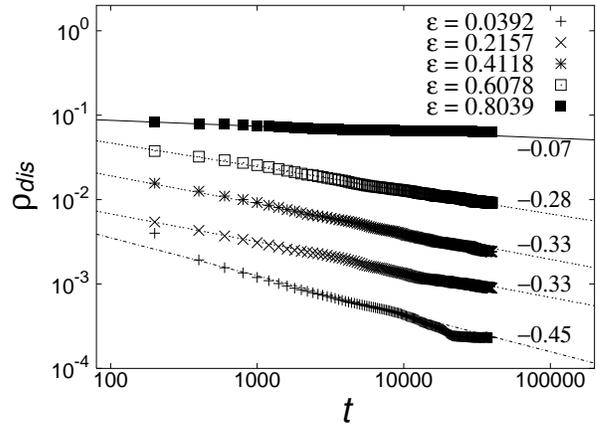}\\     
\vspace{.2in}  
\includegraphics[width=.45\textwidth]{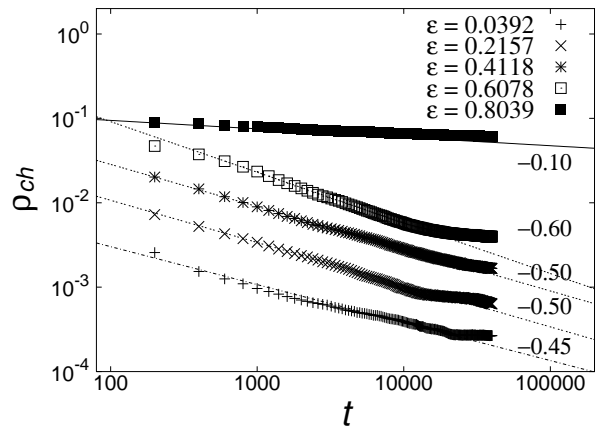}     
\caption{\label{fig:m1defect}Dislocation and chevron densities vs. time 
for Model I at different values of $\epsilon$. The curves have been 
displaced from their original position for clarity.}
\end{figure}

\begin{figure}[htp]      
\includegraphics[width=.45\textwidth]{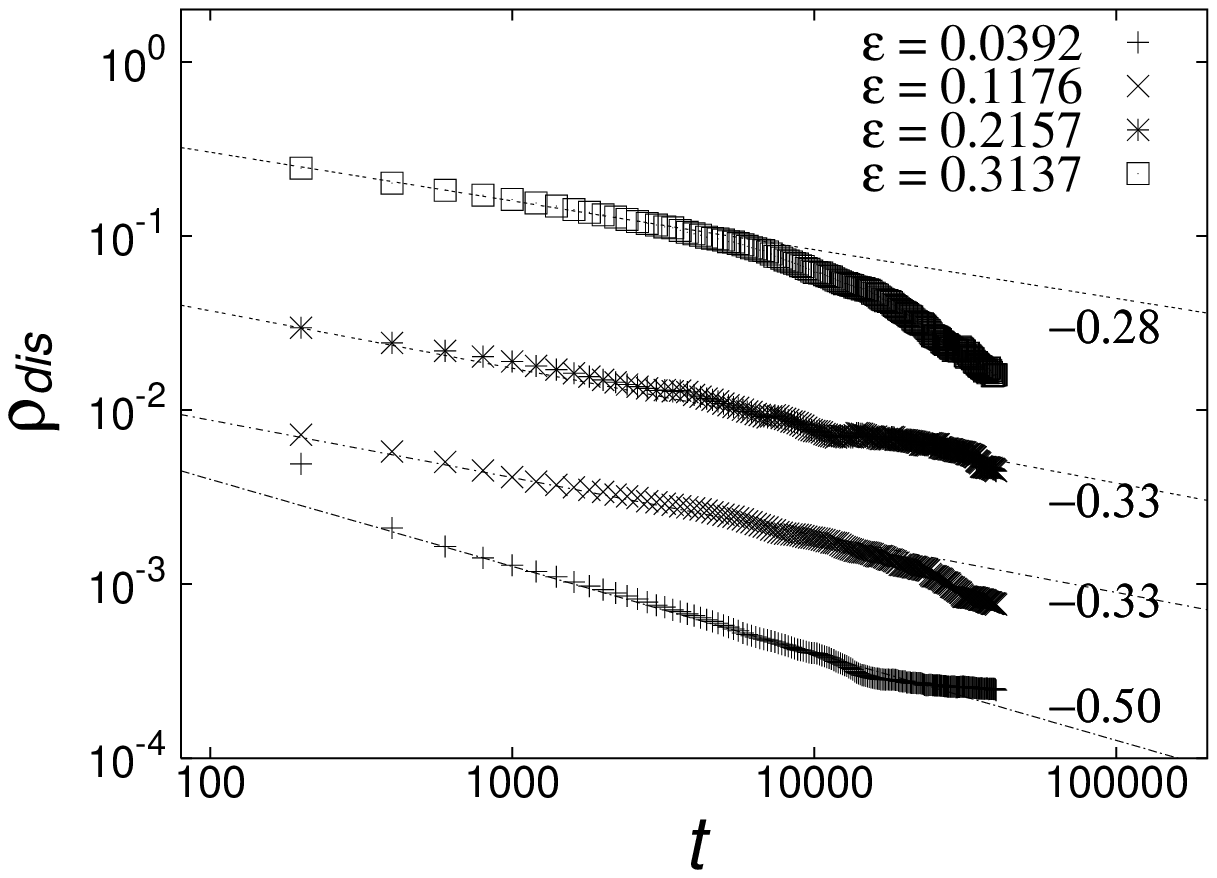}\\
\vspace{.2in}    
\includegraphics[width=.45\textwidth]{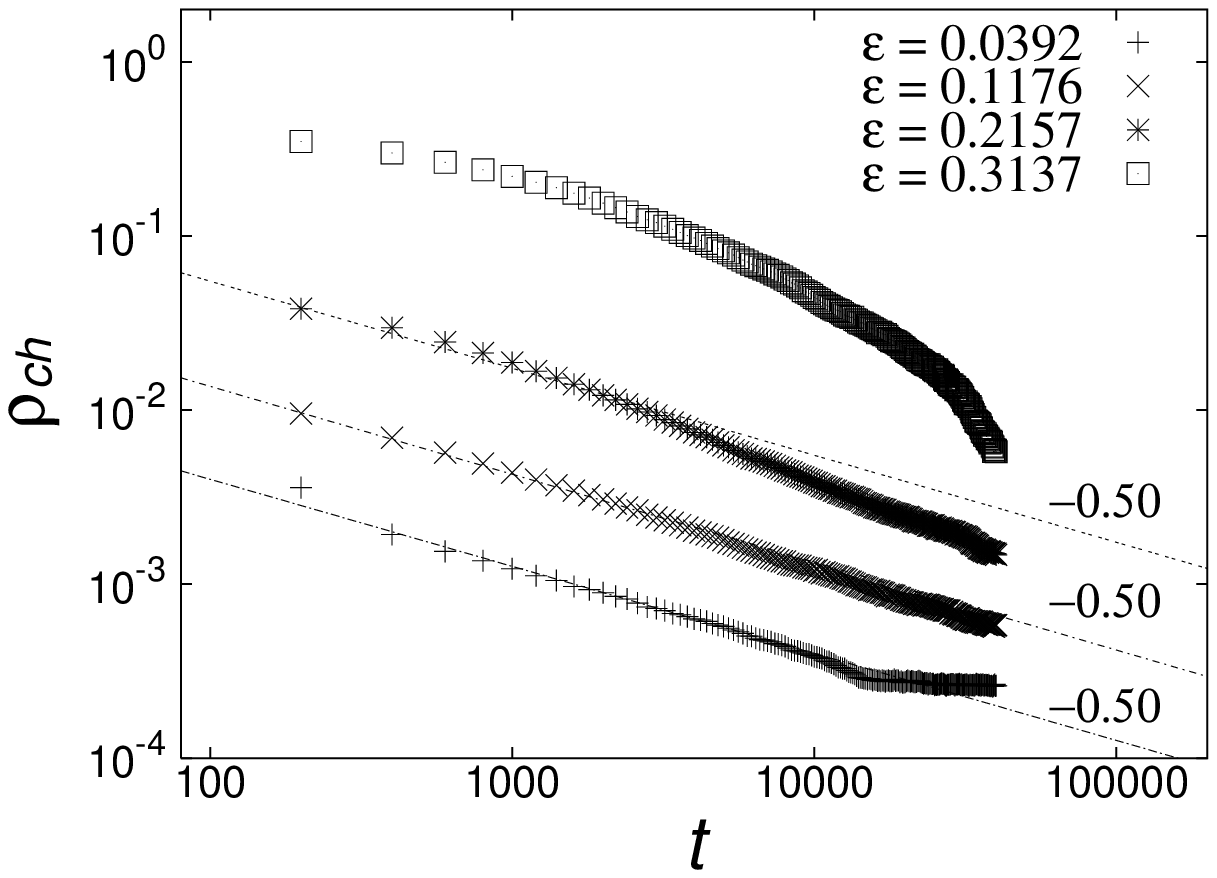}   
\caption{\label{fig:m2defect}Dislocation and chevron densities vs. 
time for Model II at different values of $\epsilon$.}
\end{figure}

\subsection{Defect densities}

The results presented above are further supported by studying the decay rate 
of defect densities. Assuming that scaling holds, the total length of 
chevron boundaries in a system of area $A$ can be estimated as 
$L_x A/(L_xL_y)\sim 1/L_y$. Similarly, the dislocation density should 
scale as $1/L_x$. Therefore, one should expect
\begin{equation}\label{correldefect}	
\rho_{dis}(t) \sim L_x(t)^{-1}, \ \rho_{ch}(t)\sim L_y(t)^{-1},
\end{equation}
or
\begin{equation}\label{eq:disloclawonset}\rho_{dis}(t) \sim t^{-1/z_{dis}}, 
\ \rho_{ch}(t) 	\sim t^{-1/z_{ch}},
\end{equation}
with the defect exponents obeying $z_{dis}=z_x$ and $z_{ch}=z_y$.

Figs. \ref{fig:m1defect} and \ref{fig:m2defect} show the 
time  evolution of $\rho_{dis}(t)$ and $\rho_{ch}(t)$ for different 
values of $\epsilon$ in both nonpotential models.

Close to onset ($\epsilon = \textrm{0.0392}$), the time evolution
of dislocation and chevron densities is well fitted by power laws with 
$1/z_{dis} \approx 1/z_{ch} \approx \textrm{0.45}$, as shown in 
Fig. \ref{fig:m1defect} (Model I) and Fig. \ref{fig:m2defect} 
(Model II). This result further suggests that defect dynamics 
is isotropic close to onset, in agreement with the arguments exposed 
below in Section \ref{sec:modelA}. Deviations from the laws 
(\ref{eq:disloclawonset}) are observed at late times ($10^4<t$) and 
are due to finite size effects (see below). 

By increasing $\epsilon$, defect dynamics becomes anisotropic and 
different laws are observed for the two types of defects. 
Chevron boundaries become quickly pinned and remain practically 
immobile whereas coarsening is dominated by dislocation 
gliding along the horizontal direction. The reduction of the length of 
a chevron boundary can occur by the motion of the slightly curved 
dislocation arrays of opposite Burgers vector located on its sides
(see Section V). 
This mechanism was also identified experimentally in electroconvection 
\cite{kamaga}. 

In an intermediate range of values of $\epsilon$, the relations 
(\ref{correldefect}) are still fulfilled, despite that chevrons boundaries
are pinned. As shown in Fig. \ref{fig:m1defect} for Model I, for 
$\epsilon = 0\textrm{.}2157$ and $\epsilon = 0\textrm{.}4118$, 
$\rho_{dis}(t)$ and $\rho_{ch}(t)$ decay as power laws with two 
different exponents, $z_{dis} \approx 3$ and $z_{ch} \approx 2$. 
These are the same values as observed in the previous section for 
$z_x$ and $z_y$, respectively. Model II leads to the same two exponents for 
$\epsilon = \textrm{0.1176}$ and $t<10^4$, see Fig. \ref{fig:m2defect}.

\begin{figure}[htp]       
\includegraphics[width=.45\textwidth]{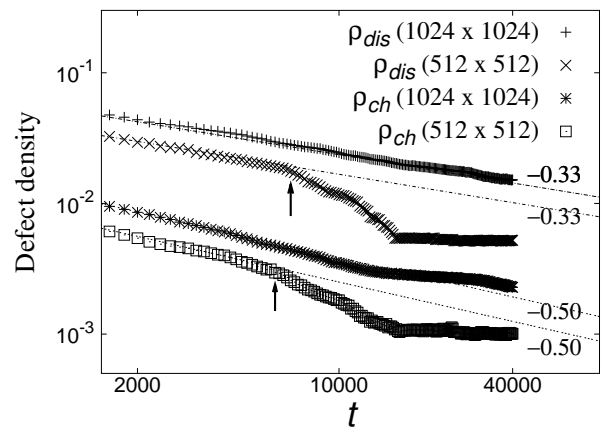}     
\caption{\label{fig:chdis}Late time evolution of the defect densities 
for Model I ($\epsilon=\textrm{0.2157}$)
for two different lattice sizes ($1024\times1024$ and $512\times512$ nodes).
The arrows indicate the beginning of finite size effects.}
\end{figure}

Fig. \ref{fig:chdis} shows 
the evolution of the defect densities obtained from two system sizes,
$1024\times1024$ and $512\times512$, with $\epsilon=0.2157$. 
As expected, corrections to scaling 
occur sooner for the smaller system (at $t \approx 7000$),
when the correlation lengths become comparable to the system size.

Note that the $t^{-1/3}$ decay rate for the dislocation density 
agrees well with the 
experimental results reported in \cite{kamaga}.
However, in this intermediate quench depth regime, our numerical results 
predict a much faster decay 
of the chevron density (as $t^{-1/2}$) than observed experimentally 
($t^{-1/5}$ \cite{kamaga}). It is worth noticing that no coarsening 
laws with exponent $1/2$ were ever observed in electroconvection
experiments.

At large quenches, dislocation gliding becomes very slow in Model I,
as shown in Fig. \ref{fig:m1defect} for $\epsilon = \textrm{0.8039}$.
For $t>2000$ 
the defect density saturate to an almost constant value, indicating 
freezing. As shown below in Section \ref{sec:nonrel}, single defect 
dynamics in Model I presents some interesting nonrelaxational 
features at large quenches, that prevent the annihilation of small 
domains.

In Model II, defect annihilation is on the contrary accelerated at large 
quenches and densities decay faster than inverse power-laws, as shown 
in Fig. \ref{fig:m2defect} for $\epsilon = \textrm{0.2157}$ and 
$\textrm{0.3137}$. These results confirm those of Section \ref{sec:correl} 
regarding the correlation lengths. Like for the other models, pinning effects 
might slow down defect motion at short times (up to $t\simeq 1000$), 
as dislocation and chevron densities decay slower 
than $t^{-1/3}$ and $t^{-1/2}$, respectively.
However, nonpotential effects dominate dislocation motion when $t>1000$ 
and the oblique stripe pattern quickly reaches fully ordered 
configurations. In Section \ref{sec:nonrel}, we show that single defect 
dynamics is quite different in Models I and II far from onset. 

Note that coarsening rates faster than a $t^{1/2}$ law for oblique stripes 
have been reported in electroconvection experiments \cite{denninback}. 
These fast coarsening 
laws were observed when the background wavenumber is small, a property also
shared by Model II (see Fig. \ref{fig:qmax}).

\subsection{Dynamical scaling}

\begin{figure}[htp]        
\includegraphics[width=.45\textwidth]{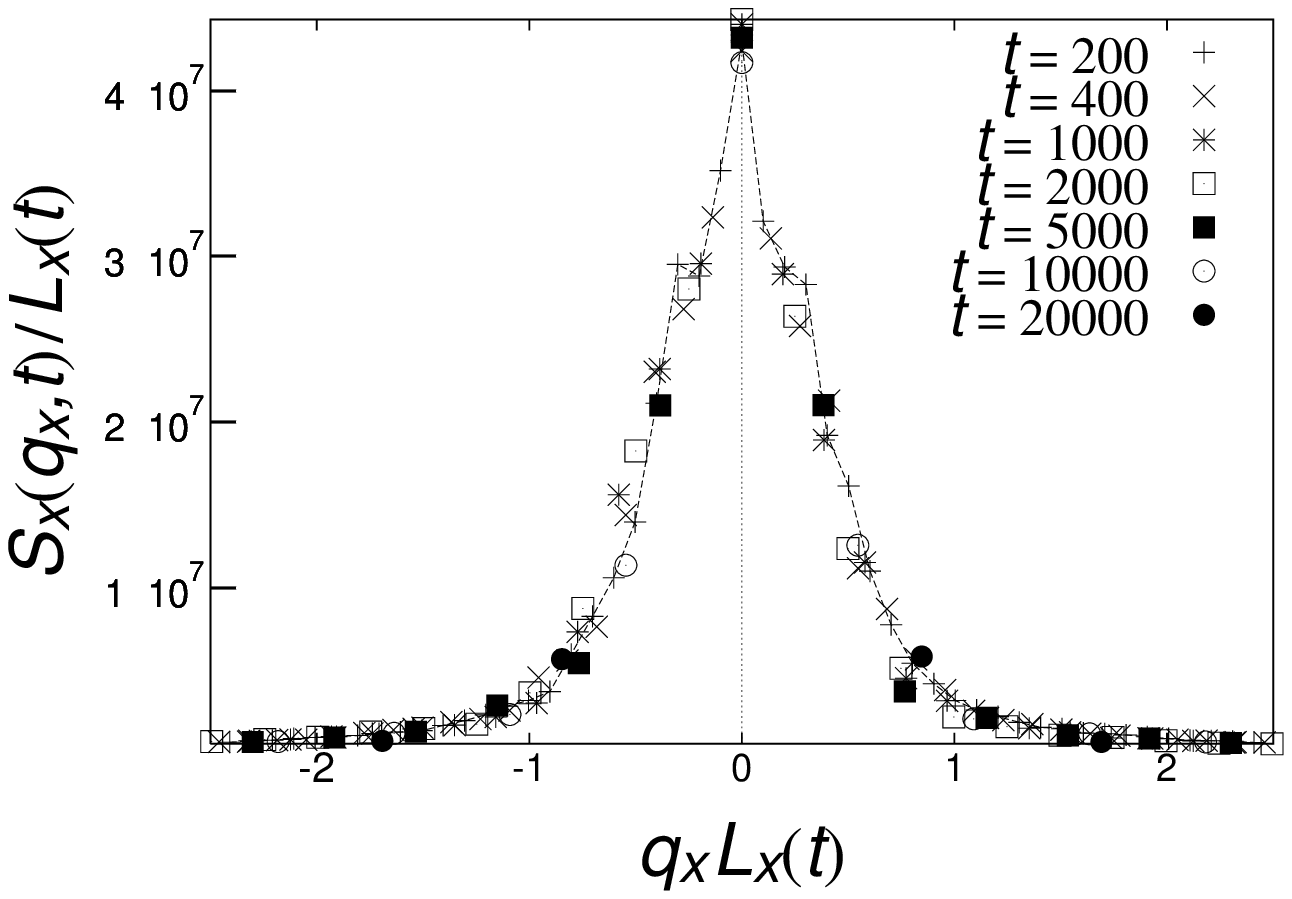}\\     
\vspace{.2in}         
\includegraphics[width=.45\textwidth]{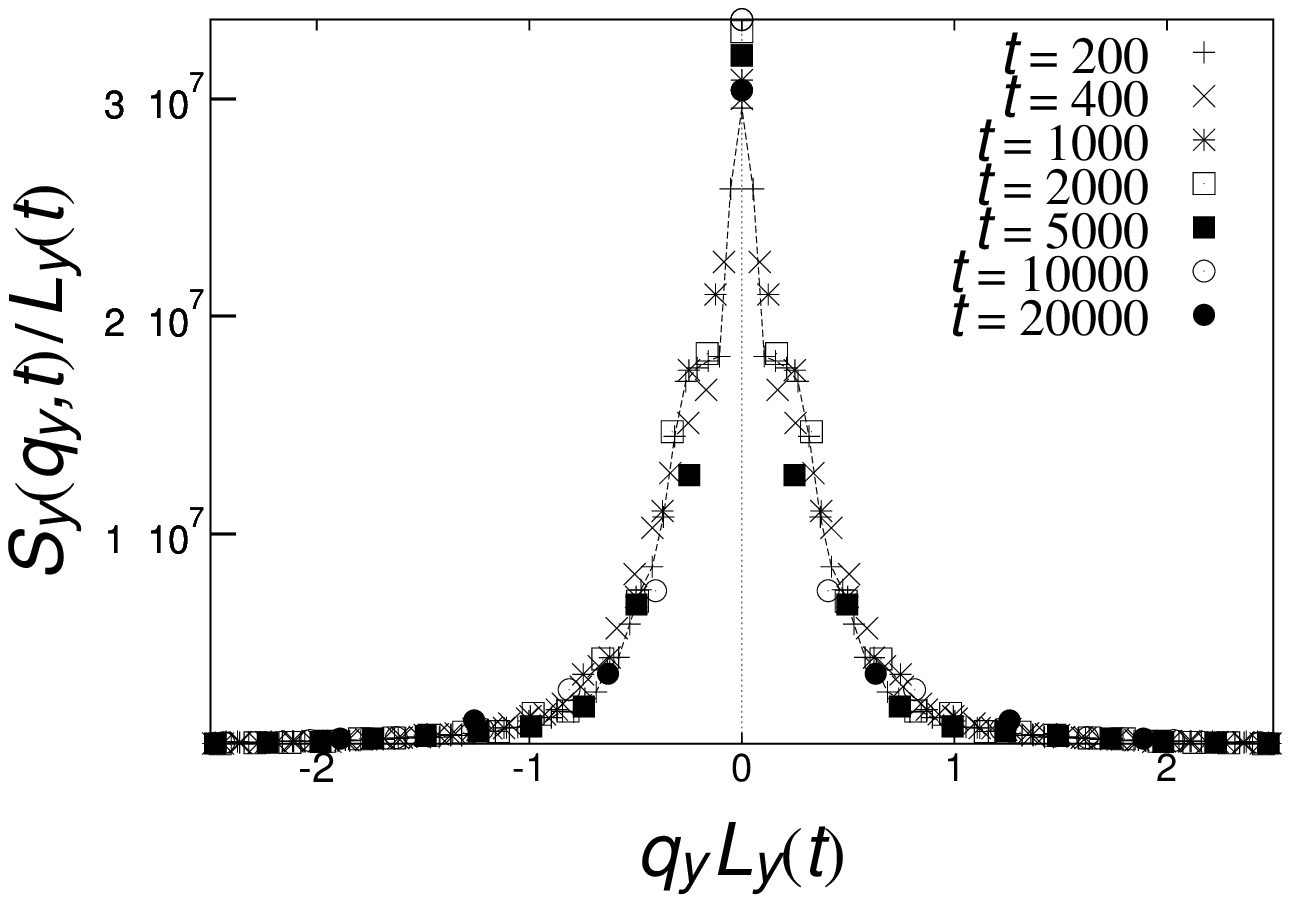}     
\caption{\label{fig:m2sxsyonset}Dynamical scaling along the $\hat{x}$ and 
$\hat{y}$  directions for Model II close to onset ($\epsilon = 
\textrm{0.0392}$).}
\end{figure}

In the case of the potential model close to onset \cite{boyer}, dynamical 
scaling properties of the structure factor were investigated along the 
directions parallel to the wavenumbers $\vec{k}_c^+$ and $\vec{k}_c^-$. 
As shown above, chevron boundaries and dislocations have very different 
mobilities further from onset. 
Therefore, it sounds more natural to test one-dimensional scaling relations 
along $\hat{x}$ and $\hat{y}$ separately. The dynamical 
scaling hypothesis implies that 
the structure factors (\ref{sx}) and (\ref{sy}) should obey the ansatz
\begin{equation}\label{ansatzsx}	
S_x(q_x,t)=L_x(t)f_x(q_x L_x(t)),
\end{equation}
\begin{equation}
\label{ansatzsy}	S_y(q_y,t)=L_y(t)f_y(q_y L_y(t)),
\end{equation}
where $L_x(t)$ and $L_y(t)$ are the characteristic length scales previously
defined through relations (\ref{lx})-(\ref{ly}); $f_x$ and $f_y$ are 
functions independent of time. 

\begin{figure}[htp]       
\includegraphics[width=.45\textwidth]{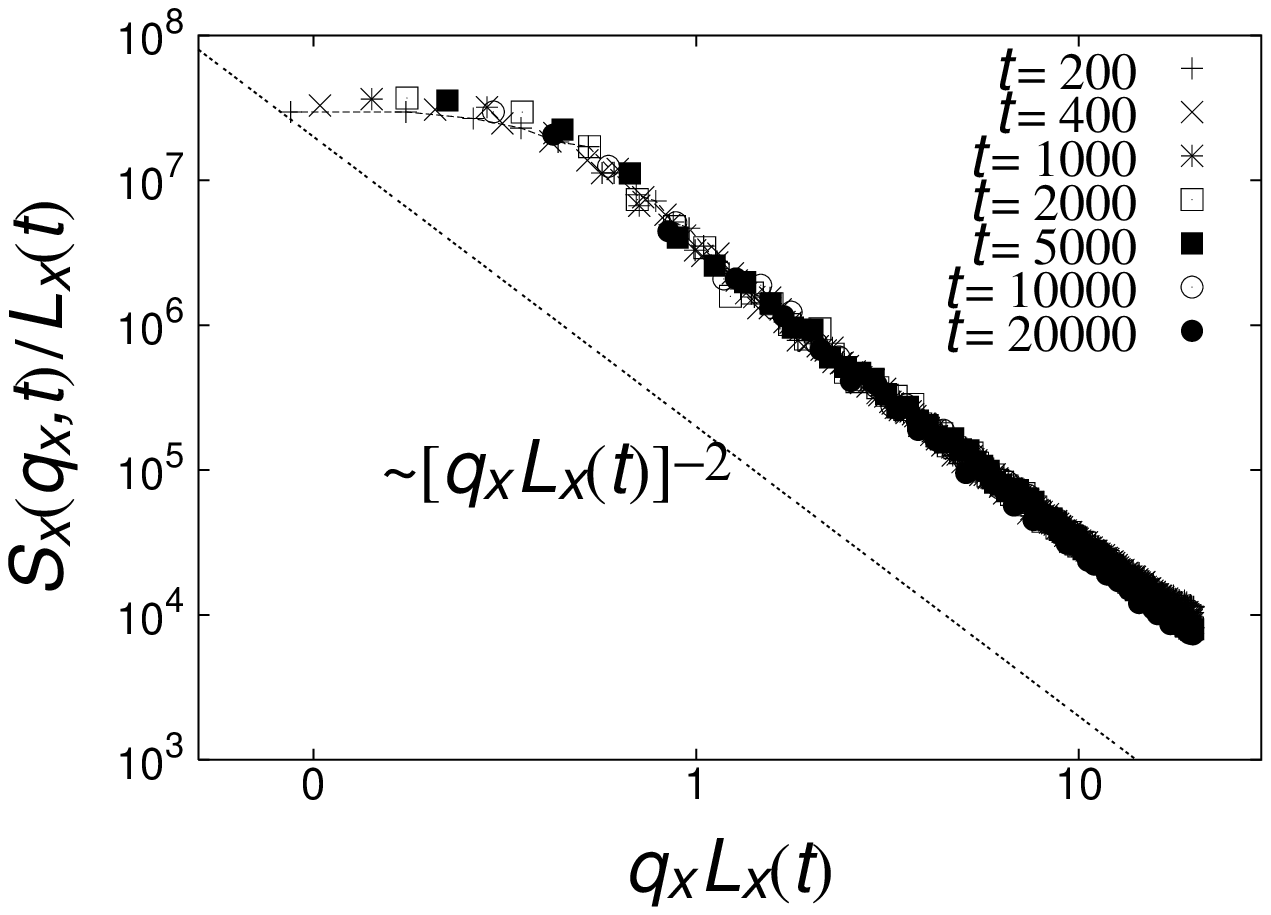}\\
\vspace{.0in}      
\includegraphics[width=.45\textwidth]{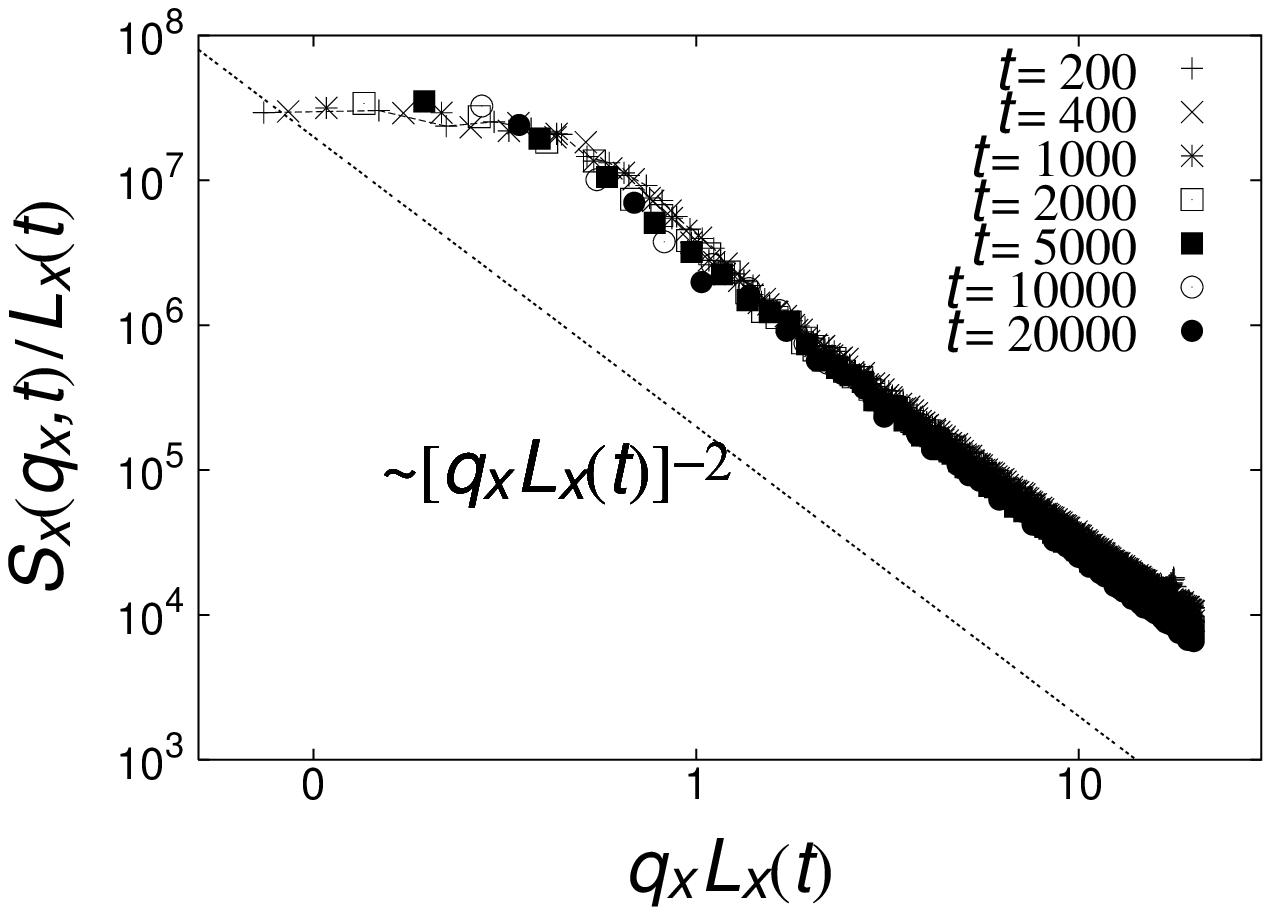}
\caption{\label{fig:m2sx}Log-log plot of $S_x(q_x,t)/L_x(t)$ as a function 
of $q_x L_x(t)$ 
for Model II: $\epsilon = \textrm{0.1176}$ (top) and      
$\epsilon = \textrm{0.2157}$ (bottom). Solid lines correspond to the Porod's 
law.}
\end{figure}

Since the ordering dynamics of the oblique striped domains close to onset 
can be approximately reduced to the dynamics of Model A (see Section 
\ref{sec:modelA}), one should expect that dynamical scaling holds along 
any direction in Fourier space. Fig. \ref{fig:m2sxsyonset} shows that 
the ansatz (\ref{ansatzsx}) and (\ref{ansatzsy}) holds over two time 
decades for Model II close to onset ($\epsilon = \textrm{0.0392}$). 
The same results are observed for the PK potential model 
and Model I (not shown). However, in the region near $k=0$, the scaling 
functions in the three models decay rapidly (see Fig. \ref{fig:m2sxsyonset}) 
and differ noticeably from that of 
Model A, which is quadratic for small $k$ \cite{otha,puri}. This unusual 
behavior may indicate that, given a random initial condition, oblique 
patterns form at early times different large-scale domain structures 
than Model A.

As $\epsilon$ is increased, isotropic dynamical scaling breaks down because 
coarsening is characterized by two growing lengths scales with
distinct exponents in the $\hat{x}$ and $\hat{y}$ directions. Nevertheless, 
Fig. \ref{fig:m2sx} (top) shows that the one-dimensional scaling relation 
(\ref{ansatzsx}) in the $\hat{x}$ direction 
(where $z_x \approx z_{dis} \approx 3$) holds over two time decades 
for Model II. In this regime, where $z_y\approx 2$,
a similar scaling behavior is also observed along $\hat{y}$ (not shown), 
for all three models. 
Deviations from dynamical scaling are observed
in the large quench depth regime, see Fig. \ref{fig:m2sx} (bottom).

We verify that the tails of the structure factors obey the 
well known Porod's Law \cite{bray}, as illustrated in Fig. \ref{fig:m2sx}. 
If the one-dimensional correlation functions obey a scaling form at short 
distances, one expects that 
$S_{x,y}(q_{x,y},t)/L_{x,y}(t) \sim 1/(q_{x,y}L_{x,y}(t))^{d+1}$ 
at large $q_{x,y}$, with $d=1$ here.

\section{Discussion}

\subsection{Mapping to Model A in an external field for low-to-moderate 
quenches}\label{sec:modelA}

\begin{figure}[htp] 
\includegraphics[width=.4\textwidth]{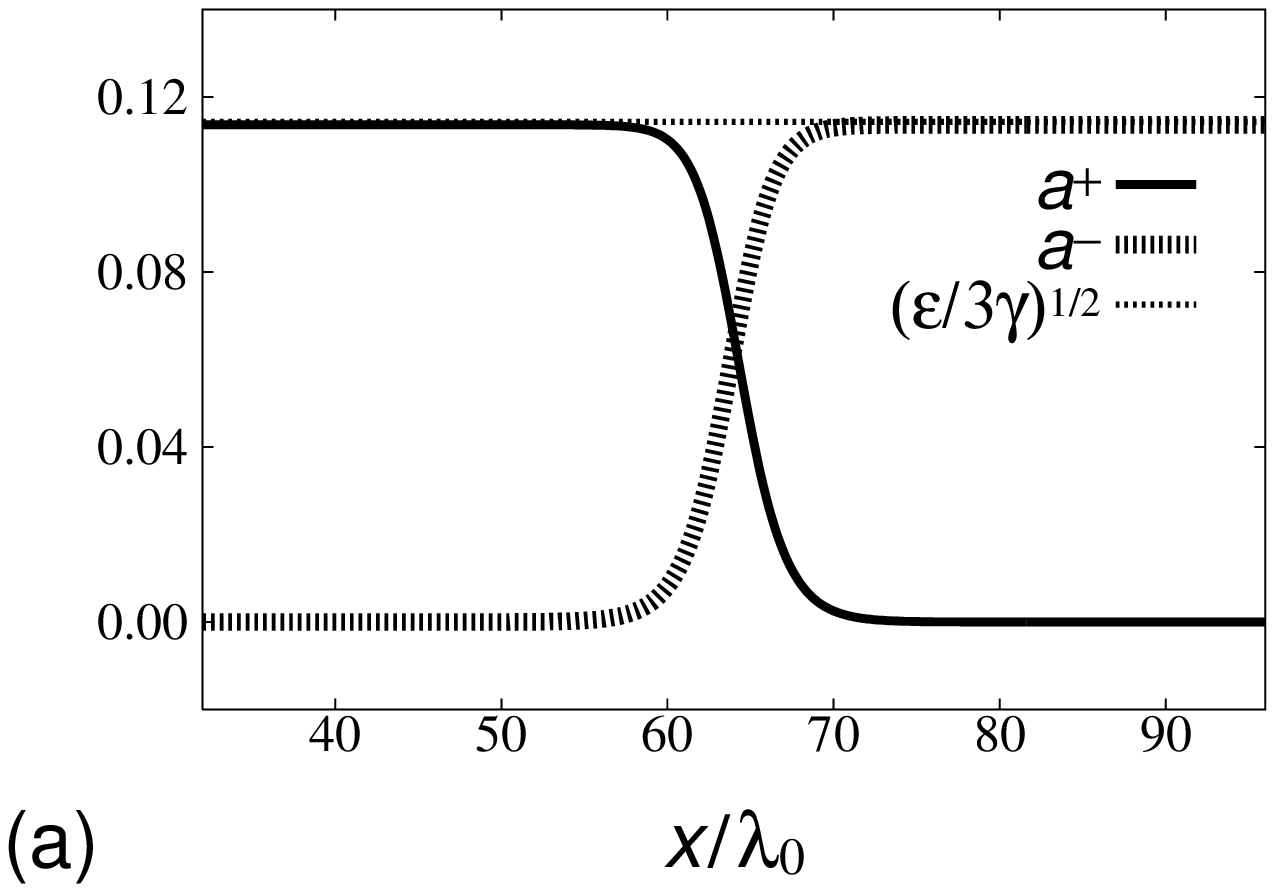} 
\vspace{.2in}         
 \includegraphics[width=.4\textwidth]{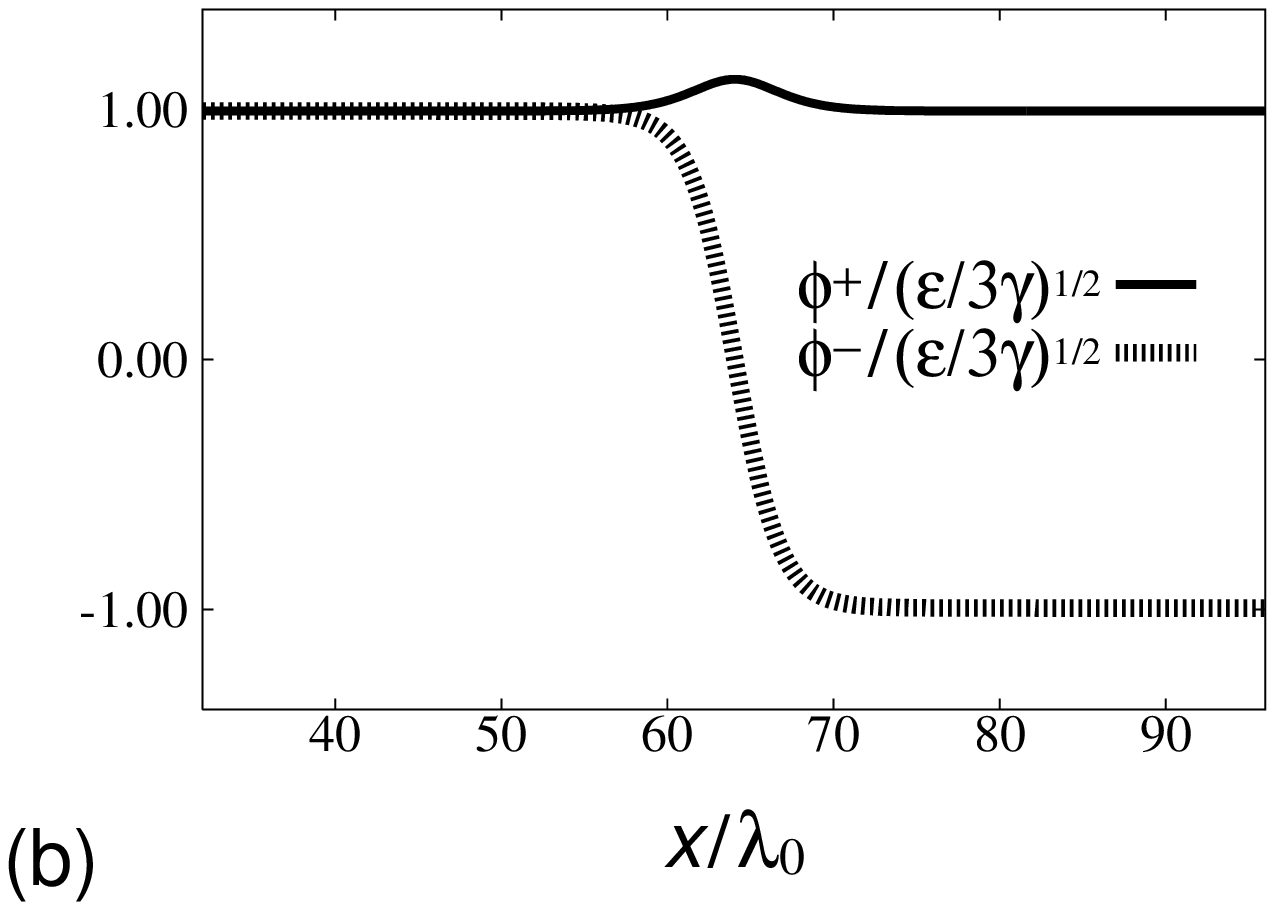}      
\caption{\label{fig:profileamp} a) Stationary numerical solutions
($a^+(x),a^-(x)$) of 
Eqs. (\ref{aniginzburg1})-(\ref{aniginzburg2}) for a vertical 
planar grain boundary, with $\epsilon = \textrm{0.0392}$,
$c=12$, $\eta = \textrm{0.5}$ and $\gamma = 1$; b) $\phi^+=a^++a^-$ 
and $\phi^-=a^+-a^-$ as a function of $x$.}
\end{figure}

The numerical results obtained {\it very close to onset} show that the
coarsening of oblique stripes is characterized by a single 
exponent, $z_{dis}\approx z_{ch}\approx z_x\approx z_y\approx 2$.
This result can be explained by the following weakly nonlinear arguments. 
The amplitude equations (\ref{aniginzburg1}) and (\ref{aniginzburg2})
can be used to describe the dynamics of grain boundaries separating zig
and zag domains close to onset ($\epsilon \ll 1$) in all models. In the 
absence of grain boundaries, the stationary uniform amplitudes in a zig 
domain are
\begin{equation}\label{steady1}	
a^+ = \sqrt{\frac{\epsilon}{3\gamma}},\ a^- = 0,
\end{equation}
and vice-versa in a zag domain. 
In the vicinity of a grain boundary, the amplitudes change continuously 
from one uniform solution to the other along the normal coordinate. 
Fig. \ref{fig:profileamp}(a) 
displays the stationary numerical profiles $a^+(x)$ and $a^-(x)$ solutions
of Eqs.(\ref{aniginzburg1})-(\ref{aniginzburg2}) close to onset 
for a vertical boundary (dislocation array). The sum $a^++a^-$ and 
difference $a^+-a^-$ are also plotted in 
Fig. \ref{fig:profileamp}(b). We observe that $a^++a^-$ remains close to 
the constant bulk value $(\epsilon/3\gamma)^{1/2}$ in the boundary region, 
while $a^+-a^-$ changes continuously from $+(\epsilon/3\gamma)^{1/2}$ 
(zig-domain) 
to $-(\epsilon/3\gamma)^{1/2}$ (zag-domain). For any polycrystalline
configuration, we define the local order parameters $\phi^+ \equiv A^++A^-$ and 
$\phi^- \equiv A^+-A^-\equiv \sqrt{\frac{\epsilon}{3\gamma}} \psi_d$, which 
re-defines the orientational order 
parameter field $\psi_d$. Based on Fig. \ref{fig:profileamp}(b),
we approximate $\phi^+$ to a constant, even in configurations 
containing many interfaces:
\begin{equation}\label{phiplus}	
\phi^+(\vec{r},t)  \simeq \sqrt{\frac{\epsilon}{3\gamma}}.
\end{equation}
To find an equation for the local orientational order 
parameter $\psi_d$, we substitute the new fields 
into (\ref{aniginzburg1})-(\ref{aniginzburg2}). By making the change 
of variables
\begin{equation}\label{scal}     
X=\left(\frac{c+2\eta-\eta^2}{c+\eta}\epsilon\right)^{1/2}x,\       
Y=\left(\frac{c+2\eta-\eta^2}{(1+c)\eta}\epsilon\right)^{1/2}y,
\end{equation}
we obtain
\begin{equation}\label{eq:psid}	
\partial_{T} \psi_d = \frac{3}{4} \psi_d + \frac{4}{k_0^2}\nabla^2 	
\psi_d - \frac{3}{4}\psi_d^3,
\end{equation}
where $T=\epsilon t$ and $\nabla^2 = \partial_X^2 +\partial_Y^2$. 
This isotropic equation is easily recast into the well-known
Model A for a non-conserved order parameter \cite{bray,halperin,allen}: 
$\partial_{t}\psi_d  =\psi_d  + \nabla^2 \psi_d  -\psi_d ^3$. Hence, as 
previously suggested \cite{boyer}, the coarsening of oblique 
stripes close to onset is isotropic (after the coordinate 
change (\ref{scal})) and driven by curvature, with a length scale 
increasing as  $t^{1/2}$. Such regime has never been observed 
experimentally in electroconvection, despite that amplitude equations 
(\ref{aniginzburg1}) and (\ref{aniginzburg2}) should {\it a priori}
describe correctly this system close enough to 
onset \cite{bodenschatz,treiber,plaut}. 

Further from onset, at intermediate values of $\epsilon$, the horizontal 
(chevron) grain boundaries become pinned in the three models considered. 
In pattern forming systems, defect pinning often arises from 
the coupling between the amplitudes varying slowly in space and the short
period modulations of the local order parameter
\cite{malomed,boyervinals2,pomeau}. This coupling generates 
\lq\lq non-adiabatic" terms in the amplitude equations, that oscillate with 
the spatial coordinates and create periodic energy barriers for defects. 
The simplest generalization of Model A that 
incorporate similar pinning effects for horizontal interfaces is 
\begin{equation}\label{externalGL}
\partial_{t}\psi_d  =\psi_d  + \nabla^2 \psi_d  -\psi_d ^3 + 
p\cos(k_py),
\end{equation}
where the periodicity of the pinning potential, of magnitude $p$, is 
{\it smaller} than the interface width ($2\pi/k_p< 1$). 
The above equation describes the ordering kinetics of a non-conserved 
scalar order parameter in a stationary, spatially modulated external field. 

As discussed in \cite{boyer}, chevron boundaries pin
at intermediate values of $\epsilon$ while dislocations are still mobile.
This effect can be explained by the fact that the magnitude of the
pinning potential of chevrons ($p$) is orders of magnitude 
higher than that of dislocation arrays. A detailed 
calculation showing this and further supporting the reduced model 
(\ref{externalGL}) will be presented elsewhere. 
Model A is recovered close to onset because pinning potentials 
tend to zero very rapidly as $\epsilon\rightarrow0$, independently of the 
interface orientation \cite{malomed,boyervinals2}. Eq. (\ref{externalGL}) 
can also be modified by replacing $p$ by a term proportional to 
$(\psi_d)^2\partial_y \psi_d$, so that the external field acts mostly 
on horizontal boundaries and not in the bulk.

\subsection{Nonrelaxational defect dynamics far from onset}\label{sec:nonrel}

\begin{figure*}[htp]      
\includegraphics[width=.23\textwidth]{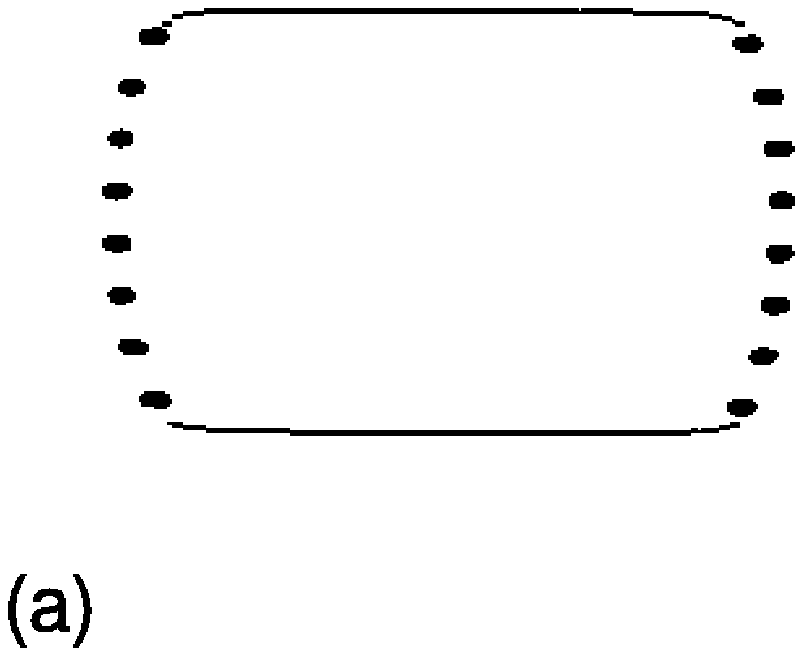}    
\hspace{.0in}          
\includegraphics[width=.23\textwidth]{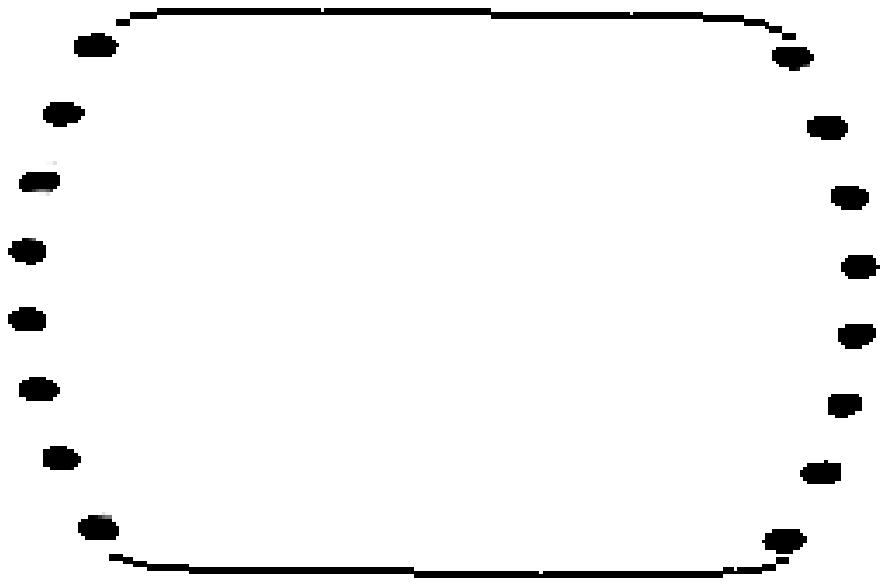}     
\hspace{.0in}         
\includegraphics[width=.23\textwidth]{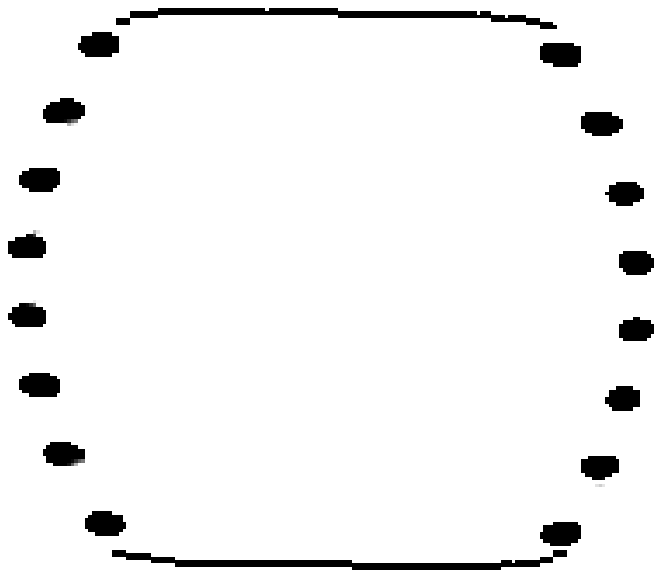}
     \hspace{.0in}        
     \includegraphics[width=.23\textwidth]{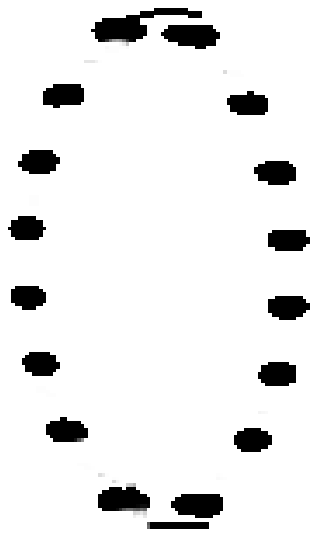}\\     
     \vspace{.0in}         
     \includegraphics[width=.23\textwidth]{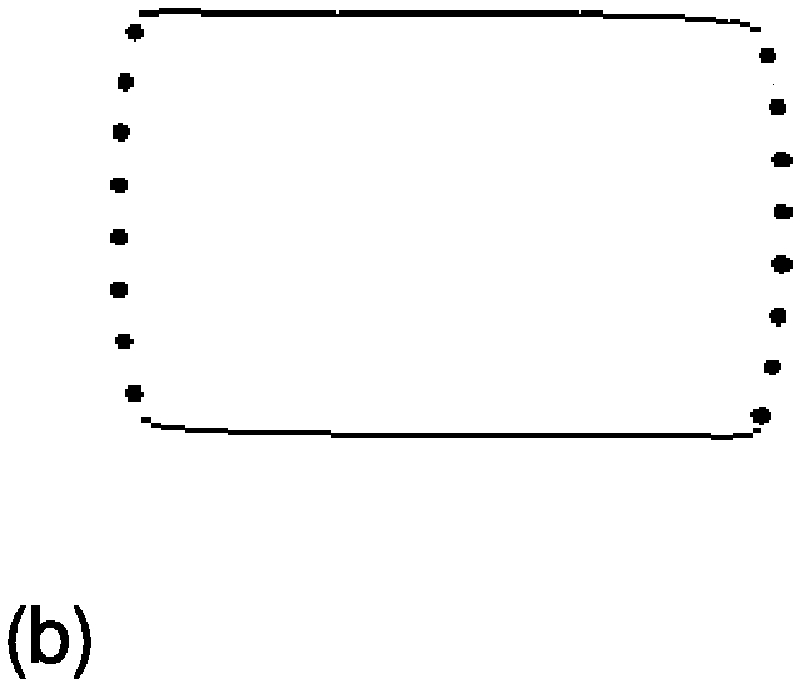}     
     \hspace{.0in}         
     \includegraphics[width=.23\textwidth]{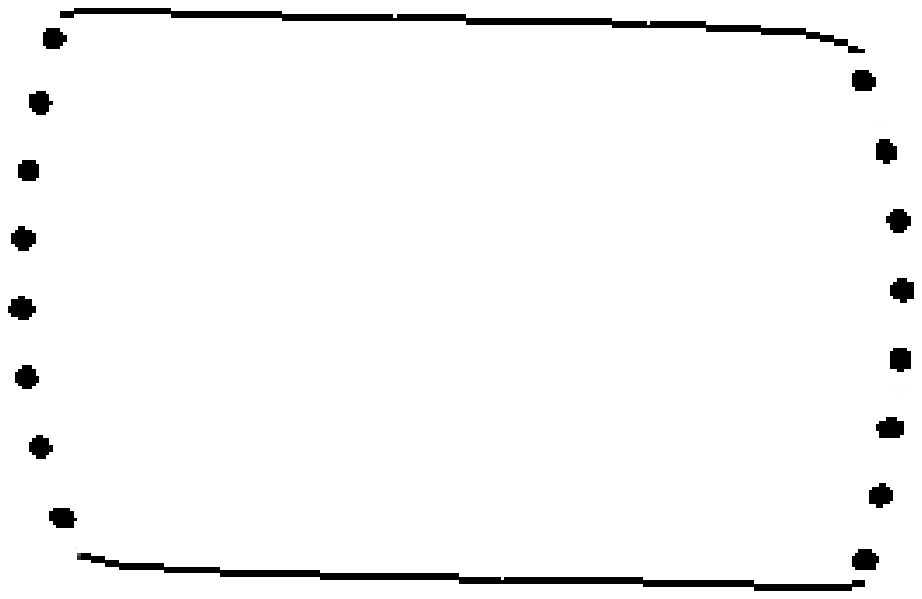}     
     \hspace{.0in}         
     \includegraphics[width=.23\textwidth]{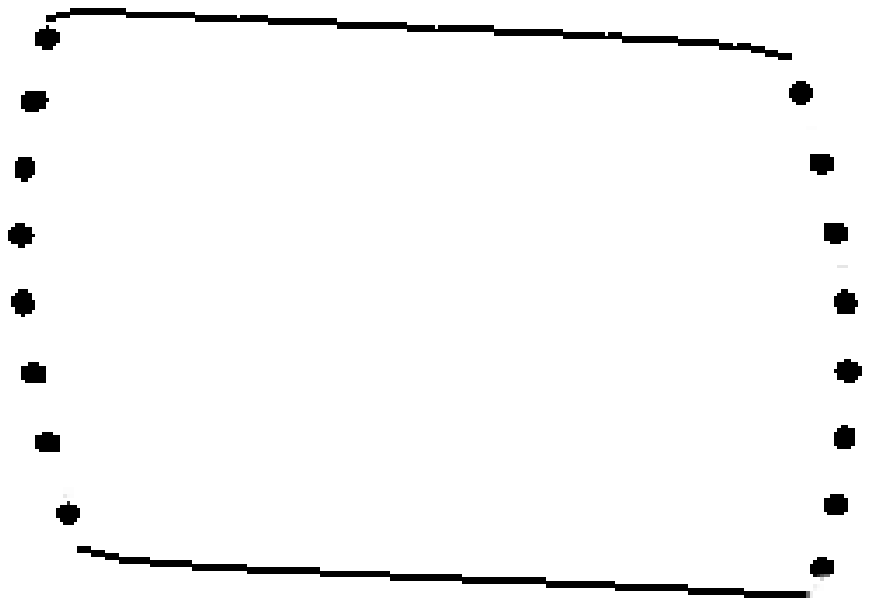}
      \hspace{.0in}         
      \includegraphics[width=.23\textwidth]{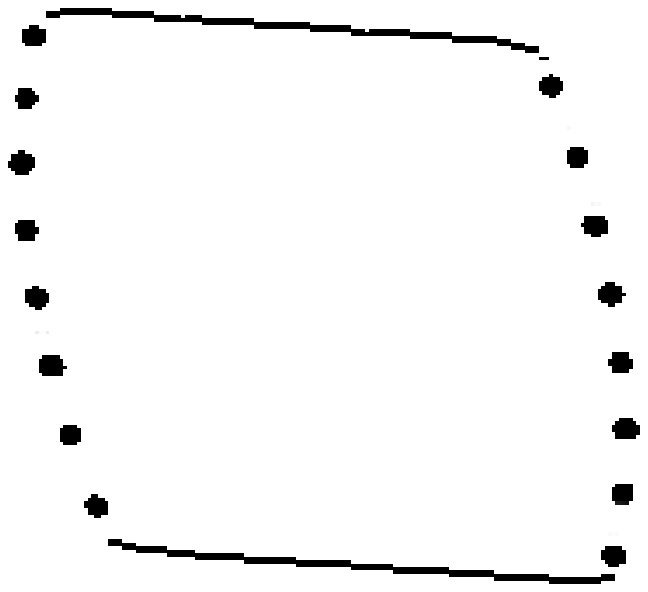}\\
      \vspace{.0in}         
      \includegraphics[width=.23\textwidth]{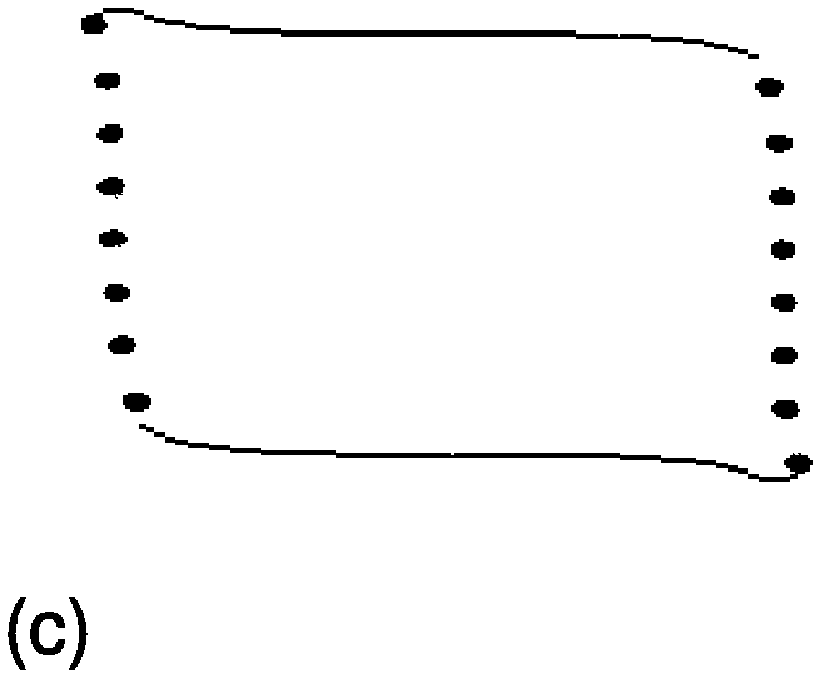} 
      \hspace{.0in}         
      \includegraphics[width=.23\textwidth]{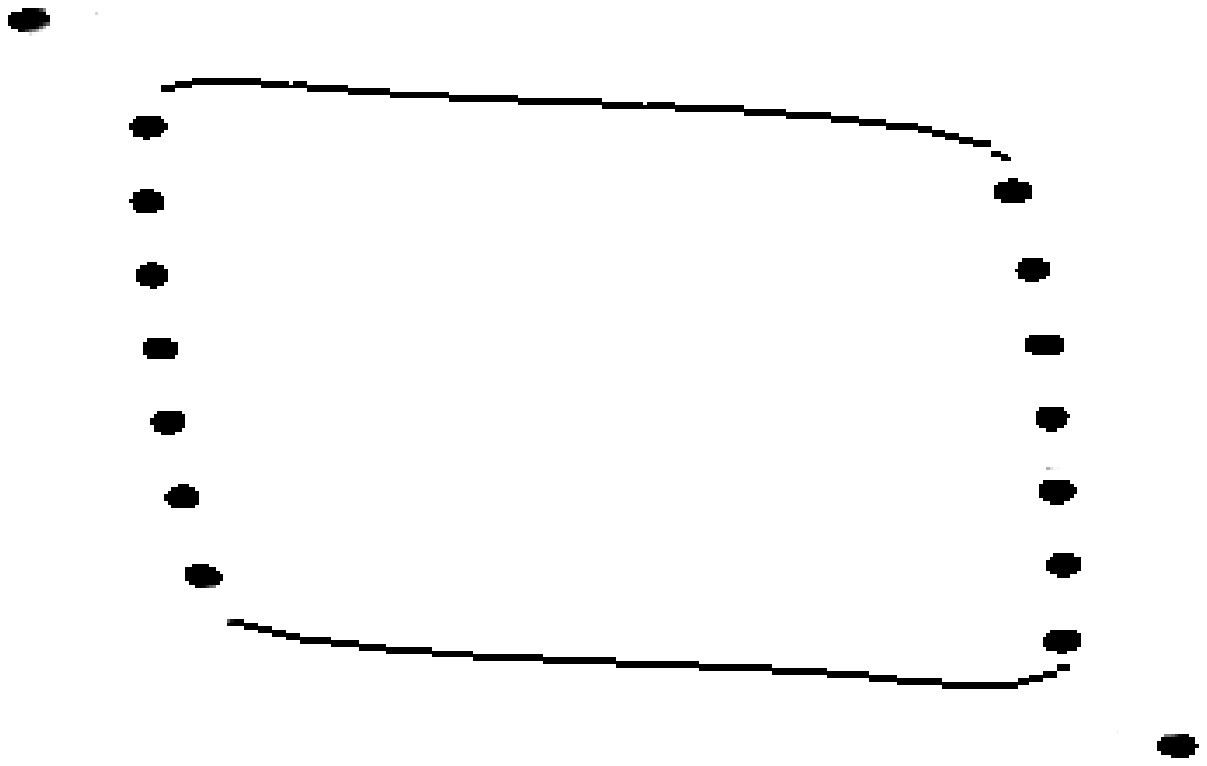}     
      \hspace{.0in}          
      \includegraphics[width=.23\textwidth]{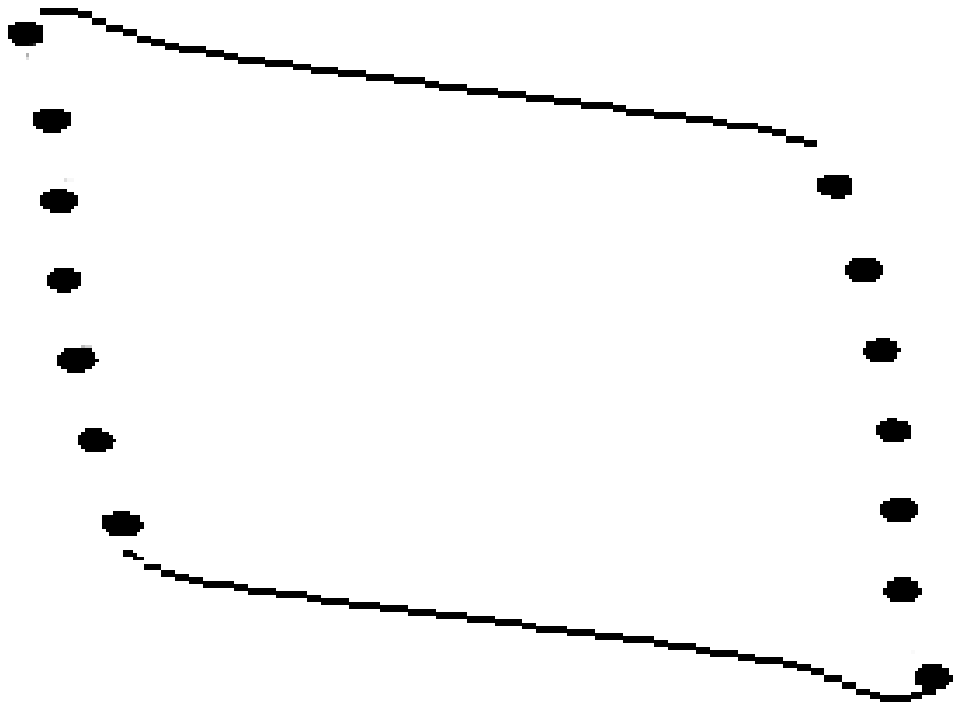}
      \hspace{.0in}         
      \includegraphics[width=.23\textwidth]{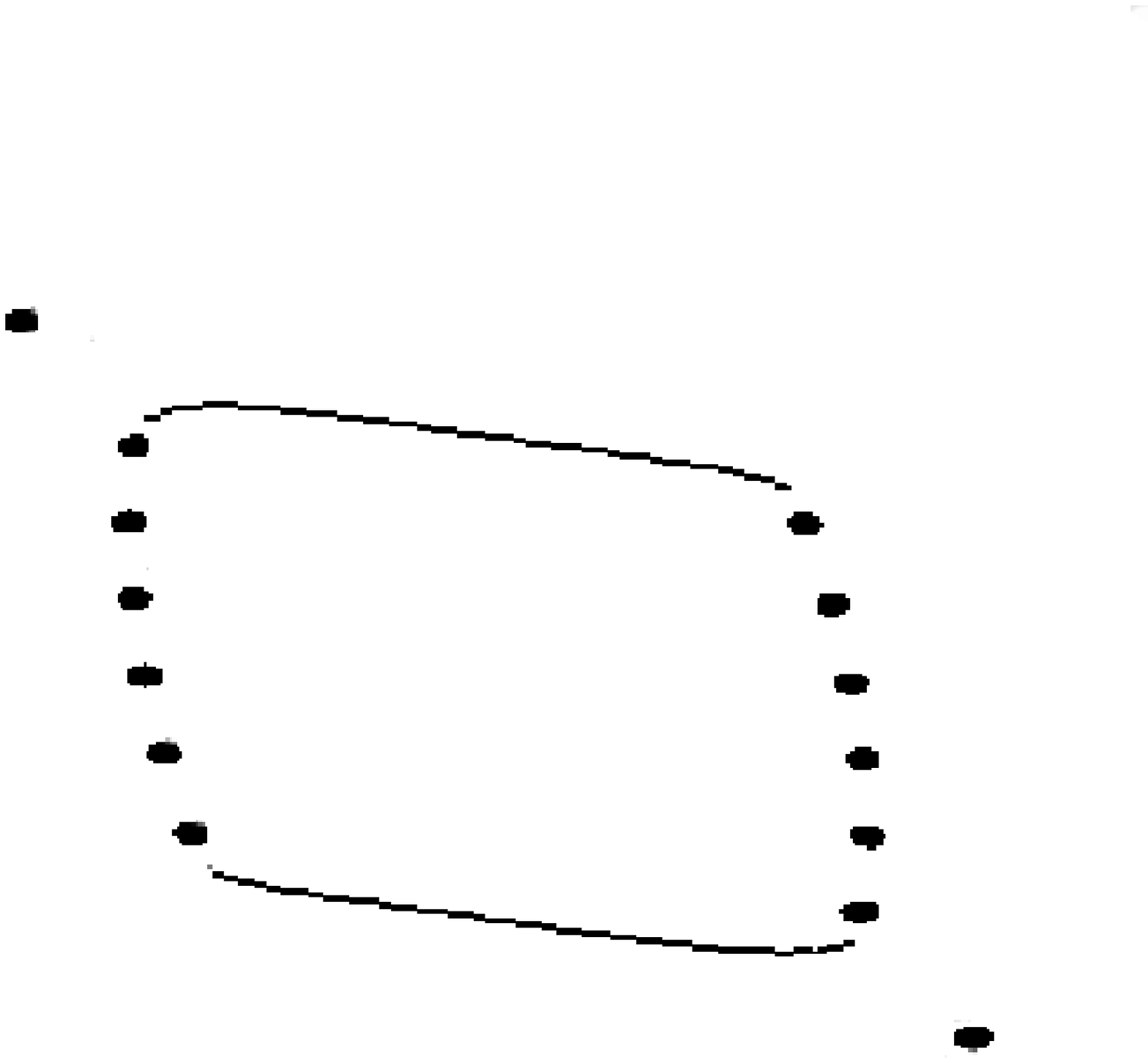}\\     
      \caption{\label{fig:singledomain}Defect field of an initially 
      rectangular zag domain surrounded by zig stripes, for Model
       I with: a) $\epsilon = \textrm{0.2157}$ at $t=200, 600, 1800$, and 
       $3600$; b) Model I with $\epsilon = \textrm{0.6078}$ at $t=200, 600, 
       1800$, and $12000$; c) Model II with 
       $\epsilon = \textrm{0.2157}$ at $t=200, 600, 1800$, and $2400$.}
       \end{figure*}

\begin{figure}[htp]    
\centering          
\includegraphics[width=.45\textwidth]{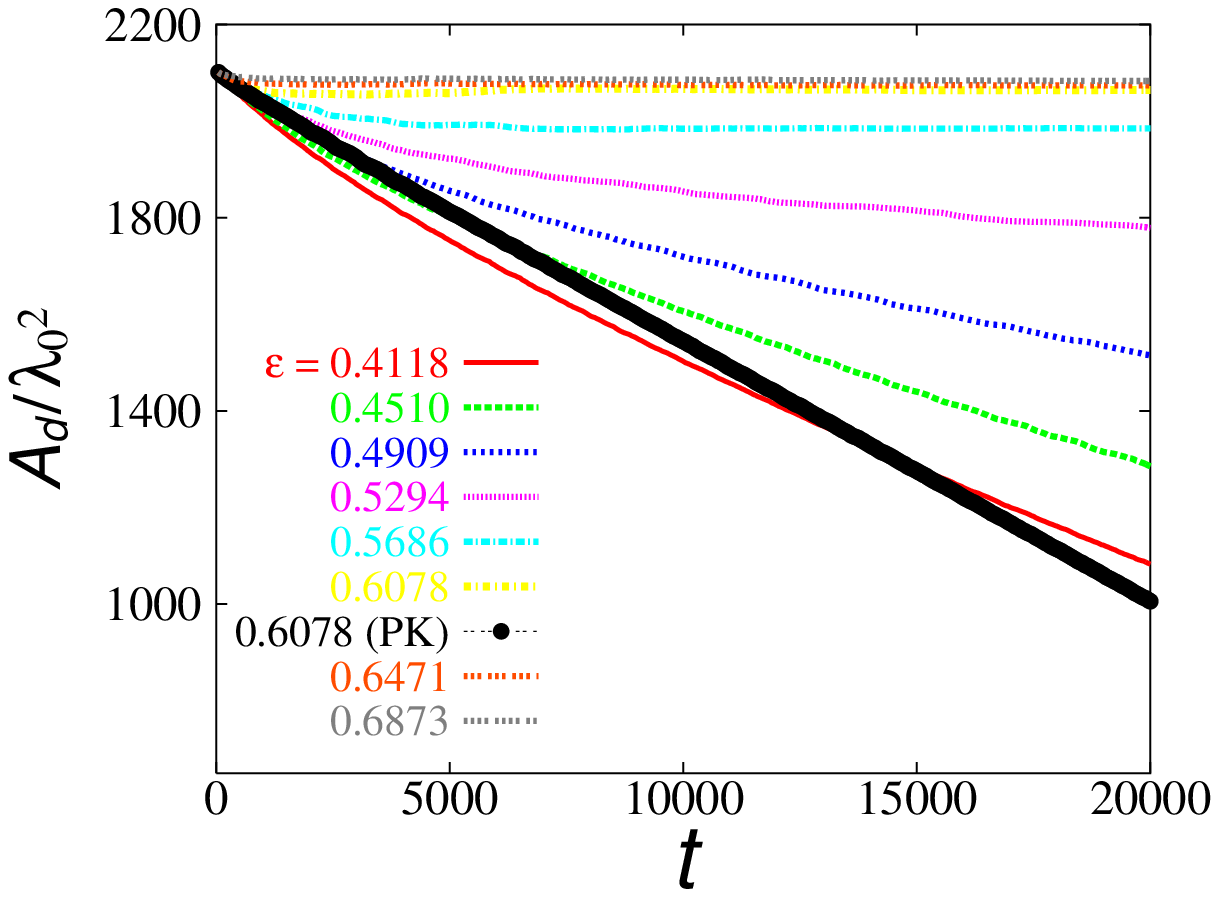}\\   
 \vspace{.0in}         
\hspace{0.5cm} \includegraphics[width=.41\textwidth]{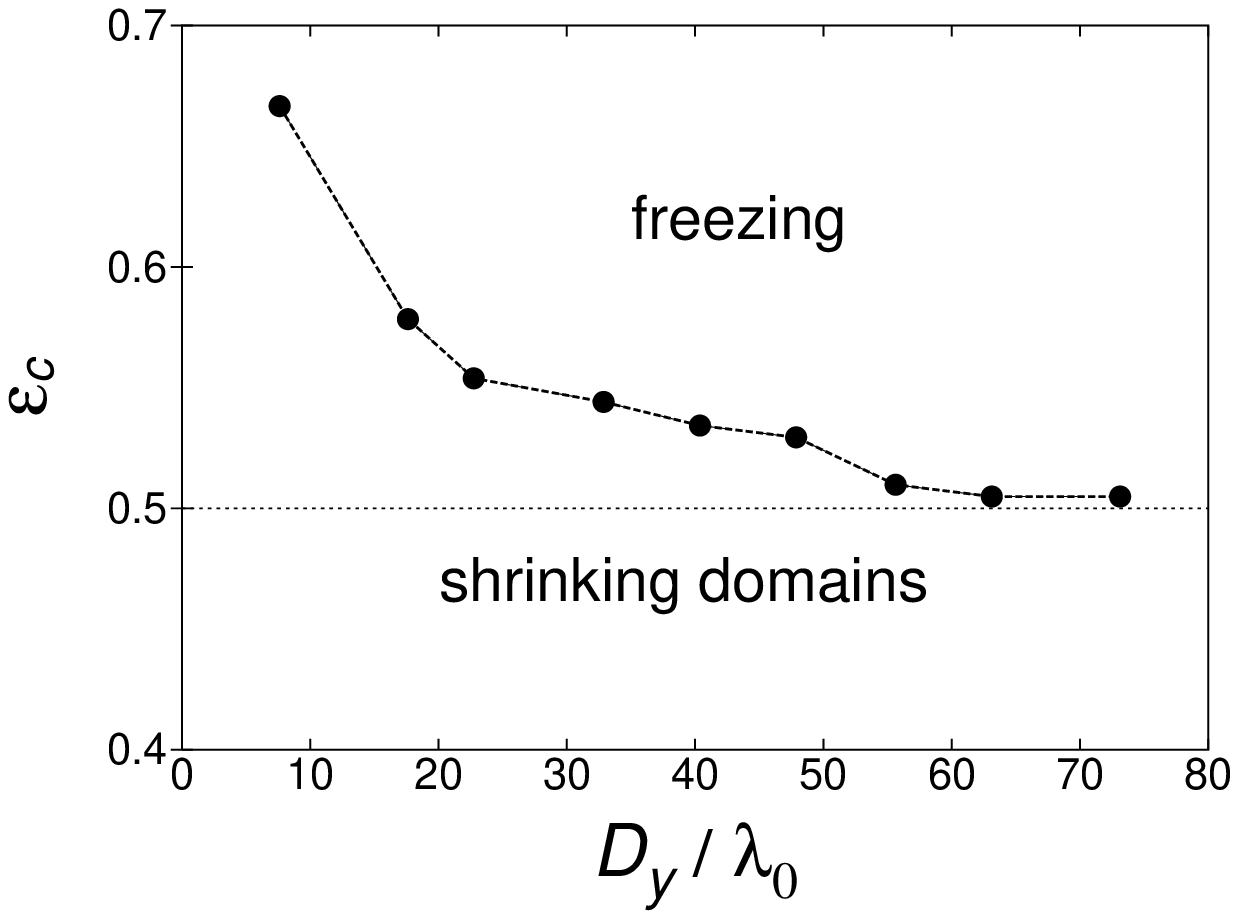}    
 \caption{\label{fig:shrink}(Color online) Upper panel: 
Time evolution of the  area of an initially 
rectangular zag domain of fixed width $D_y=32 \lambda_c$ surrounded by zig 
stripes, for Model I and different values of $\epsilon$. Lower panel: 
Critical parameter value above which a domain does not shrink, as a
function of the domain width. Below the dotted line, all domains were
observed to shrink.}
  \end{figure}

Some qualitative insights into domain coarsening for Models I and II far 
from onset can be gained by numerically studying the shrinking dynamics 
of an initially rectangular zag domain embedded in a zig surrounding. 
The background wavenumber is initially fixed to the value $k_c$. 

Fig. \ref{fig:singledomain}(a) shows the results obtained by solving
Model I with an embedded domain of vertical width $D_y=20 \lambda_c$
and $\epsilon= \textrm{0.2157}$. At this moderate quench, the dynamics is
approximately driven by the decrease of the energy (\ref{fgl}) through the
reduction of total interface length. 
As mentioned in the previous Sections, chevron boundaries are pinned and 
the reduction of their length is achieved by the motion of the curved 
lateral dislocation arrays. When colliding, dislocations of opposite
Burgers vector annihilate, as also observed experimentally \cite{kamaga}.
  
Fig. \ref{fig:singledomain}(b) 
shows the evolution of the same initial domain further from onset 
($\epsilon= \textrm{0.6078}$). In this case dislocation motion is not 
restricted to reduce the chevron length. The boundaries exhibit a 
more complex dynamics instead and the domain shape breaks its initial
axial symmetry.

To study the transition between these two dynamical regimes for Model I,
we report in the upper panel of Fig. \ref{fig:shrink} the evolution of 
the area $A_d(t)$ of an embedded domain. At moderate values of $\epsilon$, 
$A_d(t)$ continuously decreases in time at a rather constant rate as the zag 
domain shrinks. However, above a critical value, 
$\epsilon_c \approx \textrm{0.6}$ for a domain of width $D_y=32\lambda_c$, 
$A_d(t)$ tends towards a {\em finite} 
value at large times, suggesting that dislocations effectively repel 
each other. The transition between both regimes is fairly 
abrupt. This situation is reminiscent of dislocation motion modes observed 
experimentally, where walls of oppositely charged defects can approach 
within some distance and then move apart \cite{kamaga}.
The results above provide a picture of the arrested growth of order 
observed far from onset in Model I. The same single domain calculation using 
the PK potential model with $\epsilon=0.6078$, a value close to the above
$\epsilon_c$, shows that the domain completely shrinks (the bullet 
symbols of Fig.\ref{fig:shrink}). 

We therefore conclude that, unlike in
potential models, the arrested ordering kinetics in Model I is primarily 
due to nonpotential repelling effects among dislocations, and not to dislocation
pinning induced by non-adiabatic effects (although these might also be present).

The lower panel of Fig. \ref{fig:shrink} shows that, for Model I, the critical
value $\epsilon_c$ defined above depends on the domain vertical width.
Interestingly, $\epsilon_c$ seems to tend towards a constant value ($\sim 0.5$)
as $D_y\rightarrow\infty$, meaning that below this parameter value all domains 
shrink (coarsening regime). For $\epsilon>0.5$, large enough domains
do not shrink (disordered regime). This diagram suggests that Model I undergoes 
a transition to arrested coarsening as $\epsilon$ crosses a critical value.

Fig. \ref{fig:singledomain}(c) shows the evolution of a shrinking domain in 
Model II, for $\epsilon= \textrm{0.2157}$. An effective
dislocation repulsion is also observed, leading, in contrast to Model I, 
to the emission of pairs of oppositely charged dislocations. This mechanism 
gradually reduces the average domain width. For this value of $\epsilon$, 
the ordering kinetics of polycrystalline configurations is faster than a 
power law. Dislocation emission by shrinking domains at large quenches could 
be responsible for the very fast coarsening laws observed in Model II.

\section{Summary and concluding remarks}

We have studied the coarsening of two-dimensional oblique stripe phases by 
using two nonpotential Swift-Hohenberg-like equations. The results have 
been compared to previous ones obtained in the potential case \cite{boyer}.

Close to onset ($\epsilon\ll 1$), the patterns can be described in all cases 
by two coupled amplitude equations that can be reduced approximately 
to Model A for a non-conserved scalar order parameter. 
Consequently, the coarsening process close to onset is self-similar and 
domain growth is characterized by a single length scale growing as 
$t^{1/z}$ with $z \approx 2$. 

At intermediate values of $\epsilon$, a second regime is observed in both
potential and nonpotential models where chevron boundaries become pinned 
and domain growth is driven 
by the horizontal motion of curved dislocation arrays. The one-dimensional 
correlation functions along the $\hat{x}$ and $\hat{y}$ directions still 
obey dynamical scaling relations, but the corresponding correlation 
lengths grow at different rates: $L_x(t) \sim t^{1/z_x}$ and 
$L_y(t) \sim t^{1/z_y}$, with $z_x \approx 3$ and $z_y \approx 2$. 
In this regime, the system should be well described by a Model A equation 
in an external modulated field (Eq.(\ref{externalGL})). 

The nonpotential equations studied here reproduce 
qualitatively the observed selected wavenumbers in electroconvection
experiments \cite{funfschilling}. The unusual but apparently robust $t^{1/3}$ 
law for $L_x$ also agrees with experimental data \cite{kamaga}. 
However, no $1/2$ exponents were ever observed experimentally. 

When $\epsilon\sim O(1)$, a third regime can be identified, 
where defect dynamics 
differs noticeably in potential and nonpotential models. In the 
potential and nonpotential I cases, coarsening rates become very slow at 
late times, the system remaining in macroscopically disordered states. 
Whereas freezing is due to dislocation
pinning in the potential case \cite{boyer}, in Model I nonrelaxational effects 
dominate, leading to a complex dislocation dynamics (quite similar
to the experimental observations of ref. \cite{kamaga}) that prevents small 
domains to shrink. The study of the relaxation of single domains suggest
that this transition to arrested coarsening in Model I is abrupt. 
 This result is surprising, as it is generally accepted that
nonpotential effects enhance coarsening \cite{crossmeiron,yokojima,gallego}.
For Model II, nonpotential effects also
affect dislocation dynamics but lead to the opposite macroscopic behavior:
defect densities decay in time faster than a power-law and the system 
orders very quickly. 

A better understanding of the transitions between the different coarsening 
regimes far from onset requires a systematic study of the 
dynamics of shrinking domains and its role in wavenumber selection. In this 
spirit, a study of grain boundary motion in nonpotential stripe patterns 
was recently performed for isotropic systems \cite{huangvinals}. We hope that 
the study of defect motion in nonpotential systems far from onset 
will motivate new experiments. 

Finally, recent experiments \cite{griffith} studying the impact of noise 
on the coarsening rates of oblique rolls could also motivate future research.
 
\begin{acknowledgments}
This work was supported by CONACYT (Mexico) grant 40867-F.
\end{acknowledgments}


\begin{thebibliography}{9}

\bibitem{gunton} J. D. Gunton, M. San Miguel, and P. S. Sahni, 
\emph{Phase Transitions and Critical Phenomena} (Academic, New York, 1989).
	
\bibitem{bray} A. Bray, Adv. Phys. \textbf{43}, 357 (1994).

\bibitem{halperin} P. Hohenberg and P. Halperin,  Rev. Mod. Phys. 
\textbf{49}, 435 (1977).

\bibitem{brayrutenberg} A. J. Bray and A. D. Rutenberg, Phys. Rev. E 
\textbf{49}, R27 (1994).

\bibitem{bray1} A. J. Bray, Phys. Rev. E \textbf{58}, 1508 (1998).
	
\bibitem{lifshitz} I. M. Lifshitz and V. V. Slyozov, J. Phys. Chem. 
Solids \textbf{19}, 35 (1961).

\bibitem{allen} S. M. Allen and J. W. Cahn, Acta Metall \textbf{27}, 
1085 (1979).
 	
\bibitem{crosshohenberg} M. C. Cross and P. C. Hohenberg, Rev. Mod. Phys. 
\textbf{65}, 851 (1993).
	
\bibitem{bowman} C. Bowman and A. C. Newell, Rev. Mod. Phys. \textbf{70}, 
289 (1998).

\bibitem{gollub} J. P. Gollub and J. S. Langer, Rev. Mod. Phys. \textbf{71}, 
S396 (1999).

\bibitem{rabinovich} M. I. Rabinovich, A. B. Ezersky, and P. D. Weidman, 
\emph{The Dynamics of Patterns} (World Scientific, 2000).

\bibitem{elder1} K. R. Elder, J. Vi\~nals, and M. Grant, Phys. Rev. Lett. 
\textbf{68}, 3024 (1992).

\bibitem{elder2} K. R. Elder, J. Vi\~nals, and M. Grant, Phys. Rev. A 
\textbf{46}, 7618 (1992).

\bibitem{crossmeiron} M. C. Cross and D. I. Meiron, Phys. Rev. Lett. 
\textbf{75}, 2152 (1995).

\bibitem{hou} Q. Hou, S. Sasa, and N. Goldenfeld, Physica A \textbf{239}, 
219 (1997).

\bibitem{christensen} J. J. Christensen and A. J. Bray, Phys. Rev. E 
\textbf{58}, 5364 (1998).

\bibitem{boyervinals1} D. Boyer and J. Vi\~nals, Phys. Rev. E \textbf{64}, 
050101(R) (2001).

\bibitem{taneike} T. Taneike, T. Nihei, and Y. Shiwa, Phys. Lett. A 
\textbf{303}, 212 (2002).

\bibitem{boyervinals2} D. Boyer and J. Vi\~nals, Phys. Rev. E \textbf{65}, 
046119 (2002).

\bibitem{qian1} H. Qian and G. F. Mazenko, Phys. Rev. E \textbf{67}, 
036102 (2003).

\bibitem{qian2} H. Qian and G. F. Mazenko, Phys. Rev. E \textbf{68}, 021109 
(2003).

\bibitem{paul} M. R. Paul, K-H. Chiam, M. C. Cross, and P. F. Fischer, Phys. 
Rev. Lett. \textbf{93}, 064503 (2004).

\bibitem{galla} T. Galla and E. Moro, Phys. Rev. E \textbf{67}, 035101(R) (2003).

\bibitem{xu} A. Xu, G. Gonnella, A. Lamura, G. Amati, and F. Massaioli, 
Europhys. Lett. \textbf{71}, 651 (2005).

\bibitem{yokojima} Y. Yokojima and Y. Shiwa, Phys. Rev. E \textbf{65}, 
056308 (2002).

\bibitem{harrison1} C. Harrison \emph{et al.}, Science \textbf{290}, 1558 
(2000).

\bibitem{harrison2} C. Harrison \emph{et al.}, Phys. Rev. E \textbf{66}, 
011706 (2002).

\bibitem{ruiz} R. Ruiz, R.L. Sandstrom, and C.T. Black, Adv. Mater. \textbf{19},
587 (2007).

\bibitem{manneville} P. Manneville, \emph{Dissipative Structures and Weak 
Turbulence} (Academic, New York, 1990).

\bibitem{huangvinals} Z.F. Huang and J. Vi\~nals, Phys. Rev. E \textbf{75},
056202 (2007).

\bibitem{purvis} L. Purvis and M. Dennin, Phys. Rev. Lett. \textbf{86}, 
5898 (2001).

\bibitem{kamaga} C. Kamaga, F. Ibrahim, and M. Dennin, Phys. Rev. E 
\textbf{69}, 066213 (2004).
	
\bibitem{griffith} M. Griffith and M. Dennin, Phys. Rev. E \textbf{74}, 
027201 (2006).

\bibitem{boyer} D. Boyer, Phys. Rev. E \textbf{69}, 066111 (2004).
	
\bibitem{mazenko} H. Q. Qian and G. F. Mazenko, Phys. Rev. E \textbf{73}, 
036117 (2006).	

\bibitem{peschkramer} W. Pesch and L. Kramer, Z. Phys. B: Condens. Matter 
\textbf{63}, 121 (1986).

\bibitem{kramer} L. Kramer and W. Pesch, Annu. Rev. Fluid. Mech., 
\textbf{27}, 515 (1995).
	
\bibitem{buka} A. Buka and L. Kramer, \emph{Pattern Formation in 
Liquid Crystals} (Springer-Verlag, New York, 1996).

\bibitem{dennin} M. Dennin, D. S. Cannell, and G. Ahlers, Phys. Rev. E 
\textbf{57}, 638 (1998).

\bibitem{funfschilling} D. Funfschilling, B. Sammuli, and M. Dennin, 
Phys. Rev. E \textbf{67}, 016207 (2003).

\bibitem{greenside} H. S. Greenside and M. C. Cross, Phys. Rev. A 
\textbf{31}, 2492 (1985).

\bibitem{malomed} B. A. Malomed, A. A. Nepomnyashchy, and M. I. Tribelsky, 
Phys. Rev. A \textbf{42}, 7244 (1990).

\bibitem{bodenschatz} E. Bodenschatz, W. Pesch, and L. Kramer, Physica D 
\textbf{32}, 135 (1988).
	
\bibitem{treiber} M. Treiber and L. Kramer, Phys. Rev. E \textbf{58}, 
1973 (1998).
	
\bibitem{plaut} E. Plaut and R. Ribotta, Eur. Phys. J. B \textbf{5}, 
265 (1998).
	
\bibitem{chaos} M. C. Cross, D. Meiron and Y. Tu, Chaos \textbf{4}, 
607 (1994).

\bibitem{nelson} D. R. Nelson, \emph{Defects and Geometry in Condensed
Matter Physics} (Cambridge University Press, Cambridge, 2002).

\bibitem{hari} A. Hari and A. A. Nepomnyashchy, Phys. Rev. E \textbf{61}, 
4835 (2000).

\bibitem{tesauro} G. Tesauro and M. C. Cross, Phys. Rev. A \textbf{34}, 
1363 (1986).

\bibitem{denninback} C. Kamaga, D. Funfschilling and M. Dennin,
Phys. Rev. E {\bf 69}, 016308 (2004).

\bibitem{otha} T. Ohta, D. Jasnow and K. Kawasaki, Phys. Rev. Lett.
{\bf 49}, 1223 (1982).

\bibitem{puri} S. Puri and Y. Oono, Phys. Rev. A {\bf 38}, 1542 (1988).

\bibitem{pomeau} Y. Pomeau, Physica D \textbf{23}, 3 (1986).

\bibitem{gallego} R. Gallego, M. S. Miguel, and R. Toral, Phys. Rev. E 
\textbf{58}, 3125 (1998).
	
\end{thebibliography}
\end{document}